\documentclass[pmlr]{jmlr}% new name PMLR (Proceedings of Machine Learning)

\RequirePackage{graphicx}
\usepackage{booktabs}
\usepackage{multirow, multicol}
\usepackage{soul}
\usepackage{makecell}
\usepackage{longtable}% for long tables
\usepackage{listings}
\makeatletter
\def\set@curr@file#1{\def\@curr@file{#1}} %temp workaround for 2019 latex release
\makeatother
\usepackage[load-configurations=version-1]{siunitx} % newer version

\theorembodyfont{\upshape}
\theoremheaderfont{\scshape}
\theorempostheader{:}
\theoremsep{\newline}

\jmlrvolume{[VOLUME \# TBD]}
\jmlryear{2026}
\jmlrworkshop{Machine Learning for Healthcare}

\title[CBR-to-SQL Improves Healthcare Text-to-SQL]{CBR-to-SQL: Rethinking Retrieval-based Text-to-SQL using Case-based Reasoning in the Healthcare Domain}

\author{\Name{Hung Nguyen}
       \Email{hung.6.nguyen@aalto.fi}\\ 
       \addr Department of Computer Science\\
       Aalto University\\
       \AND
       \Name{Hans Moen}
       \Email{hmoen@dcm.aau.dk}\\ 
       \addr Department of Clinical Medicine\\
       Aalborg University\\
       \AND
       \Name{Pekka Marttinen}
       \Email{pekka.marttinen@aalto.fi}\\ 
       \addr Department of Computer Science\\
       Aalto University\\
} 

\begin{document}

\maketitle

\begin{abstract}
Extracting insights from Electronic Health Record (EHR) databases often requires SQL expertise, creating a barrier for clinical decision-making and research. A promising approach is to use Large Language Models (LLMs) to translate natural language questions into SQL through Retrieval-Augmented Generation (RAG), where relevant question-SQL examples are retrieved to generate new queries via few-shot learning. However, adapting this method to the medical domain is non-trivial, as effective retrieval requires examples that align with both the logical structure of the question and its referenced entities (e.g., drug names, procedure titles). Standard single-step RAG struggles to optimize both aspects simultaneously and often relies on near-exact matches to generalize effectively. This issue is especially severe in healthcare, as questions often contain noisy and inconsistent medical jargon. To address this, we present CBR-to-SQL, a framework inspired by Case-based Reasoning theory that decomposes RAG's single-step retrieval into two explicit stages: one that focuses on retrieving structurally relevant examples, and one that aligns entities with the target database schema. Evaluated on two clinical benchmarks, CBR-to-SQL achieves competitive accuracies compared to fine-tuned methods. More importantly, it demonstrates considerably higher sample efficiency and robustness than the standard RAG approach, particularly under data scarcity and retrieval perturbations.
\end{abstract}

\section{Introduction}

% Context & text-to-SQL task
Electronic Health Record (EHR) databases collect large volumes of patient data daily, including demographics, medical conditions, and diagnoses, which are critical for clinical decision-making and research. However, extracting insights from these databases often requires expertise in SQL and database schemas \citep{ziletti2024retrievalaugmentedtexttosqlgeneration}, which is challenging for non-technical users. This technical barrier can lead to inaccurate database queries, directly contributing to flawed cohort selection, overlooked patient subgroups, and ultimately unreliable clinical conclusions \citep{zella2022sqlgeneration, sauerehr2022}. Meanwhile, traditional EHR interfaces often rely on rigid, rule-based mappings, which fall short in addressing the variability of real-world queries \citep{honavar2020electronic}. 

% The technical problem
A potential solution is to utilize Large Language Models (LLMs) for text-to-SQL translation, converting natural language questions into executable SQL queries, allowing users to focus on describing the information they are seeking rather than the technical details of query construction. While generalist LLMs have demonstrated strong performance on open-domain text-to-SQL benchmarks, they struggle in specialized settings due to a lack of contextual grounding \citep{hong2025nextgenerationdatabaseinterfacessurvey}. Fine-tuning improves domain adaptation but is costly and unsuitable for dynamic environments \citep{thorpe2024dubosqldiverseretrievalaugmentedgeneration}. 

Retrieval-Augmented Generation (RAG) has gained considerable popularity as a promising alternative \citep{mohammadjafari2025naturallanguagesqlreview}, augmenting the LLM at inference time with relevant question-SQL examples retrieved from a knowledge base, allowing for few-shot generalization without requiring any parameter updates. However, effective retrieval requires examples that align with both the logical structure of the query and the specific entities it references (e.g., drug names, diagnosis names, procedure tiles). Standard RAG strategies often rely on a single-step retrieval approach, which struggles to satisfy both criteria. For instance, a retrieved example may share the exact logical structure needed to answer the question, yet be ranked low because its entities do not match those in the query, while an example whose entities align closely may be ranked high despite having an irrelevant logical structure. This challenge is especially severe in specialized domains like healthcare, where entities are noisy and often contain typos or inconsistent medical jargon \citep{wang2020texttosqlgenerationquestionanswering}. As a result, most RAG designs rely on retrieving near-exact matches to generalize \citep{ragtosql}. A common workaround is to expand the knowledge base to increase example diversity and matching chances \citep{luo2024incontextlearningretrieveddemonstrations}, but this introduces noise and computational overhead, ultimately harming performance and scalability.

% Solution
To address these limitations, we investigate classical Case-based Reasoning (CBR) theory \citep{cbroriginal}. CBR is a problem-solving paradigm that reasons by analogy through four stages: \textit{retrieving} similar past experiences, \textit{reusing} their solutions, \textit{revising} them to fit the current problem, and \textit{retaining} the results for future reference (see \appendixref{apd:cbr-cycle}). This framework maps naturally onto the text-to-SQL problem: past question-SQL pairs can be retained as reusable knowledge, retrieved by similarity, and their solutions reused and revised to answer new questions. More importantly, CBR directly addresses the core limitation of standard RAG by decomposing the reasoning problem into explicit stages, allowing structural matching and entity alignment to be optimized as independent objectives, rather than compressed into a single retrieval step.

\begin{figure*}[!h]
    \centering
    \includegraphics[width=\linewidth, keepaspectratio]{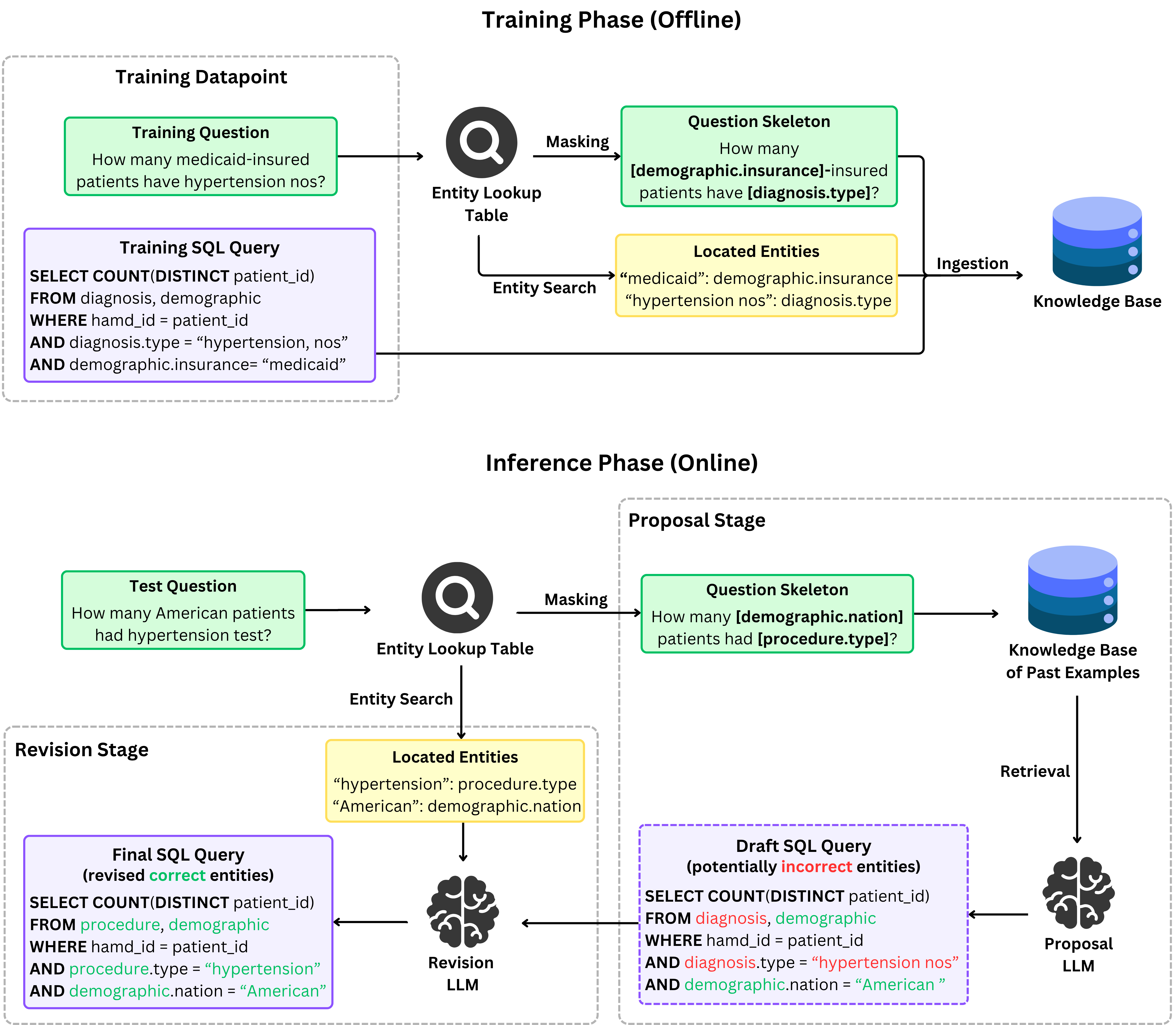}
    \caption{Overview of CBR-to-SQL's architecture.}
    \label{fig:system-architecture}
\end{figure*}

Inspired by this formulation, we propose \textit{CBR-to-SQL}, a retrieval-based text-to-SQL framework that implements the CBR workflow as two phases (\figureref{fig:system-architecture}). In the training phase, question-SQL pairs are \textit{retained} into a knowledge base for future retrieval. During the inference phase, a \textit{Proposal} stage \textit{retrieves} structurally similar question skeletons from the knowledge base and \textit{reuses} them to generate a draft SQL query. A \textit{Revision} stage then identifies the entity references in the input question, maps them to their correct locations in the database schema, and \textit{revises} the draft SQL accordingly to produce the final output.

% Contribution
Overall, our main contributions are threefold. \textbf{(1)} We propose a novel CBR-inspired formulation for medical text-to-SQL that explicitly separates logical structure retrieval from entity alignment, addressing a fundamental limitation of standard single-step RAG. \textbf{(2)} We present CBR-to-SQL, a framework that realizes this formulation through a two-stage retrieval process and a knowledge base of reusable question skeletons. Empirical results from two clinical text-to-SQL benchmarks demonstrate that this design yields considerable improvements in sample efficiency, robustness, and interpretability. \textbf{(3)} We propose a more challenging evaluation framework for retrieval-based text-to-SQL models, including an \textit{Incomplete Database} setup to evaluate performance under limited knowledge availability, and a \textit{brittleness} metric to quantify robustness to retrieval perturbations. In addition, to support reproducibility and future research, we release the code developed for this project\footnote{\href{https://anonymous.4open.science/r/CBR-to-SQL-9172/}{https://anonymous.4open.science/r/CBR-to-SQL-9172}/}.

\subsection*{Generalizable Insights about Machine Learning in the Context of Healthcare}

Reliable retrieval for clinical text-to-SQL is fundamentally limited by RAG frameworks that compress structural reasoning and entity alignment into a single step, creating systems that are brittle and unsustainable to maintain. This work identifies that the core problem is not a lack of data but a limitation of the architecture. Without a more principled retrieval strategy, naively expanding the knowledge base to improve example diversity is counterproductive, as it only introduces noise and maintenance overhead without addressing the underlying architectural limitations. Instead, we demonstrate that decomposing single-step retrieval into focused sub-tasks creates systems that are more sample-efficient and robust even in the conditions of limited training data. Our empirical results support this directly by showing that CBR-to-SQL trained on less than 10\% of available data outperforms standard RAG on the complete training dataset. Overall, we believe that the key clinical insights of this work are \textbf{(1)} demonstrating that smaller, well-curated knowledge bases with principled retrieval architectures can match or exceed the performance of data-heavy baselines, lowering the barrier for clinical institutions with limited annotation resources to deploy reliable text-to-SQL systems, and \textbf{(2)} showing that separating retrieval objectives into independent components allows for more precise error tracing and system transparency, which reduces black-box failures risks in real-world clinical deployment.

\section{Related Work}

\paragraph{Task-Specific Neural Models}

Traditional methods rely on training task-specific neural architectures end-to-end, employing specialized designs suitable for text-to-SQL \citep{sqlnet2017}. These early works primarily focus on addressing fundamental challenges such as ensuring basic SQL syntax (the \textit{order-matter} problem) using reinforcement learning \citep{zhong2017seq2sqlgeneratingstructuredqueries} or syntax-aware decoding \citep{10.5555/2969033.2969173, sqlnet2017}. In the medical domain, research focuses on introducing mechanisms to handle domain-specific terminology, such as attentive copying to align entity references in the output SQL with the schema \citep{wang2020texttosqlgenerationquestionanswering}, or encoder-as-decoder architectures to attend to entities from input questions during decoding \citep{encoderasdecoder2021}.

\paragraph{Fine-tuning}

Recent research has increasingly shifted towards fine-tuning pre-trained LLMs for text-to-SQL tasks. With their strong natural language understanding, LLMs can be adapted to complex domains such as healthcare by fine-tuning on domain-specific question-SQL pairs to internalize both the schema structure and the specialized terminology of the target domain \citep{marshan2024medt5sql}. Since many LLMs in this area were at an early stage and often lack pre-training on coding tasks, many studies also focus on developing specialized decoding modules for LLMs during fine-tuning to enhance SQL syntactic correctness \citep{scholak-etal-2021-picard, pan2021bert}.

\paragraph{In-context Learning}

As modern LLMs become more capable from pretraining on coding tasks, they can be adapted to text-to-SQL using in-context learning approaches such as RAG. Research in this field follows two parallel tracks: one is \textit{model-centric}, focusing on developing more advanced context retrieval strategies \citep{gao2023texttosqlempoweredlargelanguage, shi-etal-2025-gen}, the other is \textit{data-centric}, aiming to enhance context representation through techniques like knowledge graphs \citep{shen2025magesqlenhancingincontextlearning, li2024scalable}. In the medical domain, however, retrieval-based models remain relatively underexplored. A notable exception is \citet{ziletti-dambrosi-2024-retrieval}, which, similar to CBR-to-SQL, combines retrieval with entity alignment to handle specialized medical terminology in EHR text-to-SQL. However, their entity alignment module is restricted to clinical ontology codes, while this component in CBR-to-SQL generalizes to any natural language entity mention and has a broader role: allowing structured retrieval via entity tagging and entity revision in the final query. Furthermore, \citeauthor{ziletti-dambrosi-2024-retrieval} position medical terminology alignment as the core challenge, while we treat it as a symptom of a deeper architectural limitation with single-step RAG, allowing CBR-to-SQL to target a broader class of clinical text-to-SQL problems.

\section{Methods}

As illustrated in \figureref{fig:system-architecture}, CBR-to-SQL operates in two phases: an offline training phase that constructs a knowledge base from question-SQL pairs, and an online inference phase that utilizes it to generate SQL queries for new questions.

During the training phase, each question is transformed into a \textit{question skeleton}. We first extract entity mentions from training questions, resolve them against an \textit{Entity Lookup Table} to pinpoint their schema locations, then replace these entities with their corresponding locations as semantic tags (e.g., \texttt{[demographic.insurance]}, \texttt{[diagnosis.type]}). This masking process strips away noisy details from training questions, allowing retrieval to focus on the abstract structure of the problem and obtain more logically aligned examples. The resulting skeletons, along with their associated entities and SQL queries, are stored as examples in a knowledge base to be reused during inference. It is worth noting that this ``training" phase involves no parameter updates, but it is purely a preprocessing step.

The inference phase then proceeds in two stages. The first stage, \textit{Proposal}, focuses on structural alignment. The input question is masked using the same procedure to produce its skeleton, which is used to retrieve structurally similar examples from the knowledge base. These examples are then utilized to generate a draft SQL query that captures the correct logical structure. However, since retrieval only prioritizes structural similarity, this draft may contain incorrect entity references inherited from the retrieved examples. The \textit{Revision} stage addresses this by querying the Entity Lookup Table to identify the correct schema locations for the entities mentioned in the input question, and substituting them into the draft to produce the final executable SQL query.

\subsection{Entity Lookup Table}

The main goal of the lookup table is to translate natural language entity references into schema-specific values. It functions as a ``dictionary of synonyms" to support entity alignment in both training and inference phases.

\paragraph{Lookup Table Construction}

The lookup table is constructed offline by extracting every unique value from columns of the target EHR database, along with their corresponding locations (column, table). Each value is then indexed in a vector database, with its schema location stored as metadata. For the embeddings task, we use a specialized medical model rather than a generalist one to accurately represent domain-specific terminologies.

As indexing every unique value in the database to the lookup table would be impractical, we derive a process to target only columns that capture meaningful entities. Specifically, we retain columns that represent name-driven concepts that are likely to be referenced in natural language in varied or imprecise ways, such as medication names or procedure titles, rather than ID-like or numerical fields. To this end, we apply a two-step column filtering pipeline: a rule-based filter first excludes clearly non-textual or ID-like columns using regular expressions, followed by an LLM judge that scores the semantic richness of values based on samples from each column. The detailed process is described in \appendixref{apd:entity-categories}. This pipeline is fully automatic and can be applied zero-shot to any database schema.

\paragraph{Entity Retrieval}
\label{para:entity-retrieval}

Given an entity extracted from a natural language question, the lookup table retrieves its schema-specific matches using a two-step process. First, a semantic search using a medical embedding model and cosine similarity retrieves the top 100 candidates, prioritizing recall by capturing synonyms and paraphrases. A re-ranking step then improves precision by scoring candidates based on their Levenshtein distance to the query entity, prioritizing close syntactic matches. The final candidate list is restricted to the top 5 matches, with each including the entity value and its corresponding location. This re-ranking step is a task-specific design choice motivated by the nature of the benchmark datasets used in this study, where user queries often contain minor spelling errors or small deviations from actual entity names, making syntactic similarity particularly valuable. In scenarios where users lack domain knowledge and describe concepts loosely, such as ``stomachache" instead of ``abdominal pain", a purely semantic retrieval approach may be more suitable.

\subsection{Offline Training Phase}

\paragraph{Question Masking}
\label{para:entity-masking}

Given a dataset of question-SQL pairs, we apply a masking procedure to transform each natural language question into an abstract \textit{question skeleton}. To achieve this, an LLM first performs entity extraction on the question, identifying text spans and assigning them to a set of general semantic categories (e.g., procedure, diagnosis, patient). Each extracted entity is then resolved against the lookup table to retrieve a set of candidate schema matches, from which an LLM selects the most appropriate and links it to its schema location. The entity is then replaced with its schema location in the format \texttt{[TABLE\_NAME].[COLUMN\_NAME]} as its semantic tag. If no match is found, the entity retains its initial semantic category as a fallback tag.

Question masking has been shown to be an effective strategy for removing noisy information prior to retrieval \citep{gao2023texttosqlempoweredlargelanguage}, which motivates our masking approach. However, most existing techniques replace entities with generic category labels such as \texttt{[DISEASE]} or \texttt{[DRUG]}, which discard information about how entities relate to the underlying database schema. Instead, we utilize table and column names as tags, providing a schema-aligned representation that abstracts away noise while preserving each entity's structural role. For instance, mentions of ``heart attack" and ``myocardial infarction" in different questions both map to the same \texttt{[diagnoses.long\_title]} tag, allowing structural matching across semantically equivalent but lexically different queries.

Although LLM-based entity labeling may introduce errors due to limited medical expertise, perfect labeling is not the goal. Even approximate masking improves retrieval by shifting the focus from noisy specialized entities to the underlying query structure. For example, a question about ``oldest patients with diabetes" mistagged as ``oldest patients with \texttt{[PROCEDURE]}" can still retrieve a structurally similar example about ``oldest patients with \texttt{[DIAGNOSIS]}", as both share the same logical pattern of finding the oldest patients filtered by a medical attribute. This robustness makes the approach practical even with imperfect entity labeling. 

\paragraph{Knowledge Base Ingestion}

Once the training question-SQL pairs are masked, the resulting question skeletons, their extracted entities, and corresponding SQL queries are ingested into a vector database, indexed by dense embeddings of the masked questions. Since masking removes domain-specific medical terminology from the questions, leaving only general language patterns, a generalist embedding model is sufficient for indexing, keeping the system lightweight. Overall, this masking-based representation allows CBR-to-SQL to focus solely on retrieving similar problem patterns, while being a simpler alternative than structural abstraction approaches such as syntax trees \citep{Awasthi_Chakrabarti_Sarawagi_2023} or knowledge graphs \citep{shen2025magesqlenhancingincontextlearning}, which require extensive schema engineering to construct.

\subsection{Online Inference Phase}

\paragraph{Proposal Stage}

The Proposal stage aims to generate a draft SQL query that captures the correct logical structure of the SQL problem. For a new input question, we first apply the same masking procedure described in \sectionref{para:entity-masking} to produce its question skeleton. Example retrieval is then formulated as a Dense Passage Retrieval (DPR) problem \citep{karpukhin-etal-2020-dense}, retrieving the top-$k$ (with $k=5$) most similar skeletons from the knowledge base via nearest neighbor search using cosine similarity. An LLM conditioned on the retrieved examples then generates a draft SQL query via few-shot learning, reusing their logical structure. Since the retrieved examples are masked, the LLM receives only structural signals about the query, allowing it to prioritize resolving the SQL structure. However, this requires it to predict specific entity values and their schema locations directly from the question, which may be incorrect.

\paragraph{Revision Stage}

The Revision stage corrects the entity references in the draft SQL query by aligning them to the target schema. For each entity extracted from the input question, we perform a top-$k$ ($k=5$) search over the lookup table as described in \sectionref{para:entity-retrieval} to retrieve candidate schema matches. An LLM then selects the correct match for each entity and fills it into the draft SQL. The model is provided with the input question, database schema, and candidate entities to allow context-aware disambiguation. For instance, if an entity has similar matches in both the PROCEDURE and DIAGNOSIS tables, the LLM can use contextual signals to determine the correct schema location. The final output is a fully executable SQL query aligned with the target schema. 

\section{Experiments}
\label{sec:experiments}

\subsection{Datasets}

\paragraph{MIMICSQL} We conduct experiments using the clinical text-to-SQL dataset MIMICSQL \citep{wang2020texttosqlgenerationquestionanswering} and the underlying EHR database MIMIC-III \citep{mimic-iii}. MIMICSQL contains 10,000 machine-generated question-SQL pairs modified by human annotators to introduce real-world clinical challenges like ambiguous terminology, abbreviations, and typos. Further statistics are provided in \appendixref{apd:mimicsql-statistics}. The dataset is partitioned into training, validation, and test sets in an 8:1:1 ratio, where the training split is used for knowledge base construction and the test split for evaluation. MIMICSQL serves as our primary benchmark, as it directly targets the core challenge CBR-to-SQL aims to solve: handling ambiguous medical terminology and varied natural language formulations.

\paragraph{EHRSQL} While MIMICSQL questions feature diverse paraphrasing and controlled typos, most SQL queries in this dataset remain simple to moderate in difficulty, limiting assessment of model performance on deeply nested subqueries or complex logic. To address this, we conduct additional experiments using the EHRSQL benchmark \citep{lee-etal-2024-overview} and the MIMIC-IV Clinical Database Demo \citep{PhysioNet-mimic-iv-demo-2.2}. This dataset introduces two key challenges: deeply nested SQL queries and impossible questions that require the model to abstain. Notably, EHRSQL is explicitly designed to evaluate reliability through abstention on unanswerable queries, a dimension that is not the focus of this paper. Consequently, direct comparisons on execution accuracy should be interpreted with the caveat that most competing methods on this benchmark are optimized for answerability detection, whereas CBR-to-SQL prioritizes SQL generation accuracy. More detailed dataset statistics will be provided in \appendixref{apd:ehrsql-statistics}. The training set of EHRSQL (5,124 datapoints) is utilized for knowledge base ingestion, and the test split (1,167 datapoints) for evaluation. 

\subsection{Evaluation Setups}

We construct two evaluation setups that differ in how training data are ingested into the knowledge base, allowing us to assess CBR-to-SQL and its baseline under both high-resource and data-scarce conditions. Together, these two setups are applied to the MIMICSQL and EHRSQL dataset, producing four main experiments in total.

\paragraph{Complete Database (CDB)} The CDB setup includes the full training set in the knowledge base, similar to how standard text-to-SQL models are trained. It facilitates direct comparison with prior work that used the complete MIMICSQL and EHRSQL datasets for training and evaluation. Furthermore, CDB provides high example diversity for retrieval models, while also introducing noise from overlapping examples. This serves as a testbed to benchmark CBR-to-SQL against RAG in a high-resource setting, evaluating their ability to extract relevant examples among noisy distractors. 

\paragraph{Incomplete Database (IDB)} To stress-test robustness and sample efficiency, we introduce IDB as a more challenging evaluation setup. It simulates data scarcity conditions, where each remaining example in the knowledge base is nearly structurally unique, preventing models from relying on diverse task demonstrations as in CDB.

Prior approaches often produce incomplete data using global random dropping \citep{thai2022cbrikbcasebasedreasoningapproach, saxena-etal-2020-improving}, which risks removing the only example of a given query structure, introducing coverage gaps in the training data that require multiple runs to stabilize. To address this, our method performs stratified sampling over query structures, producing a minimal dataset while approximating the original structural coverage. To achieve this, we utilize a two-step process: SQL masking to reveal the underlying problem pattern, and clustering to aggressively reduce the SQL skeletons to a nearly unique set. This approach allows retrieval-based systems to operate in a limited data setting while preserving the core structural pattern coverage in the original training data for generalization.

First, we apply a masking procedure using regular expressions to remove entity values from SQL queries, resulting in abstract skeletons that retain column names, operators, and clause structures. These masked SQL queries are then encoded using a generalist embedding model and clustered into groups, with each group representing a distinct problem pattern. From each cluster, we retain a single random sample, while all outliers are preserved. Overall, this reduces the original training sets from 8,000 to 774 examples for MIMICSQL, and from 5,000 to 492 examples for EHRSQL, retaining only the most representative examples. Further details on the process are given in \appendixref{apd:idb-construction}.

\subsection{Metrics}

We adopt the standard text-to-SQL evaluation metrics, Execution Accuracy and Logical Form Accuracy, from MIMICSQL's paper, and introduce an additional metric, \textit{brittleness}, to stress-test generalization robustness. 

\paragraph{Execution Accuracy} $Acc_{\text{EX}}$ measures the proportion of generated queries that produce correct results when executed, defined as $Acc_{\text{EX}} = N_{\text{EX}} / N$, where $N_{\text{EX}}$ is the number of queries whose execution results match those of the ground-truth queries and $N$ is the total number of queries. However, a limitation of this metric is that it can misjudge when a SQL query accidentally produces the correct result, even without aligning to the intended logical structure.

\paragraph{Logical Form Accuracy} $Acc_{\text{LF}}$ measures the proportion of generated queries that exactly match the ground-truth SQL string, defined as $Acc_{\text{LF}} = N_{\text{LF}} / N$, where $N_{\text{LF}}$ is the number of exact matches. As it requires precise syntactic and structural alignment, $Acc_{\text{LF}}$ provides a stricter and more informative measure of generalization than $Acc_{\text{EX}}$ alone. Accordingly, a large gap between $Acc_{\text{EX}}$ and $Acc_{\text{LF}}$ is indicative of a model that frequently arrives at correct results through structurally inconsistent queries, which is a form of overfitting to the training SQL queries rather than generalizing the logic.

\paragraph{Brittleness} To further assess generalization robustness, we propose \textit{brittleness} ($\Delta_{\text{brittle}}$), a metric that quantifies a model's performance drop when top-ranked retrieved examples are removed. Given a ranked list of $k$ retrieved examples $\{c_1, c_2, \ldots, c_k\}$ where $c_1$ is the top-ranked example, we apply stochastic dropout with rank-based probabilities defined as:
\begin{align}
p_i= \begin{cases}
        p_{\text{top}} \cdot \dfrac{k - i}{k - 1} & \text{if } i \leq k, \\
        0 & \text{otherwise},
    \end{cases}
\end{align}
where $p_{\text{top}} \in [0, 1]$ is the maximum dropout probability applied to the top-ranked example ($i=1$). In this paper, $p_{\text{top}}=1$ such that for $k=5$, the dropout probabilities from rank 1 to 5 are $\{1, 0.75, 0.5, 0.25, 0\}$, respectively, following a linearly decaying rate. Linear decay is selected as it enforces a monotonic relationship between rank and importance without introducing unnecessary complexity.

Brittleness is then defined as $\Delta_{\text{brittle}} = Acc_{\text{original}} - Acc_{\text{drop}}$, where $Acc$ can be either $Acc_{\text{EX}}$ or $Acc_{\text{LF}}$. A high $\Delta_{\text{brittle}}$ indicates strong reliance on top-ranked examples and fragility, whereas a low value reflects more robust, sample-efficient generalization. 

\subsection{Baselines}

To establish a retrieval-based baseline for CBR-to-SQL, we construct \textit{RAG-to-SQL}, which represents the standard RAG approach. This model serves as the primary comparison baseline for CBR-to-SQL on both MIMICSQL and EHRSQL. It operates by performing a single retrieval step to obtain relevant question–SQL pairs as in-context examples, then generating SQL queries via an LLM with few-shot learning (\appendixref{apd:ragsql-architecture}).

For MIMICSQL, we additionally compare against recent fine-tuning methods evaluated on this benchmark. This includes \cite{CHEN2026111800}'s method, which constructs a knowledge graph modeling EHR schema and complex question-schema reasoning to inform LLM fine-tuning, and MedTS \citep{pan2021bert}, which fine-tunes a BERT encoder and applies a grammar-constrained LSTM decoder to generate syntactically correct SQL trees. Task-specific neural network methods from prior MIMICSQL experiments of \cite{wang2020texttosqlgenerationquestionanswering} are also reported. This includes TREQS \citep{wang2020texttosqlgenerationquestionanswering}, which uses attentive copying to handle medical entity references; Coarse2Fine \citep{mekala-etal-2021-coarse2fine}, which generates abstract query sketches before filling in details; SQLNet \citep{xu2017sqlnetgeneratingstructuredqueries}, which leverages syntax-aware decoding for unordered WHERE clauses; and PtrGen \citep{lukovnikov2018translatingnaturallanguagesql}, a pointer-generator model effective for low-frequency or out-of-vocabulary terms.

For the EHRSQL benchmark, we evaluate our method against participating systems from the EHRSQL 2024 shared task, following the official results reported in \citet{lee-etal-2024-overview}. PLUQ \citep{jo-etal-2024-lg} utilizes self-training with pseudo-labeled unanswerable questions filtered by token entropy and execution, then retrains to enhance question answerability detection. PromptMind \citep{gundabathula-kolar-2024-promptmind} proposes an ensemble of fine-tuned LLMs with unanimous voting for abstention and domain-specific example retrieval to augment LLM context. ProbGate \citep{kim-etal-2024-probgate} employs a fine-tuned ChatGPT model with log-probability confidence filtering and execution-based error detection to identify and abstain from unanswerable questions. KU-DMIS \citep{kim-etal-2024-ku} fine-tunes GPT-3.5-turbo on templated question formats with detailed schema prompts, generating pseudo question–SQL pairs to align with test distribution and using SQL prediction self-consistency for abstention. AIRI NLP \citep{somov-etal-2024-airi} demonstrates a lightweight approach using logistic regression for answerability detection and a fine-tuned T5-3B for SQL generation. LTRC-IIITH \citep{thomas-etal-2024-ltrc} utilizes SQLCoder-7b-2 for both answerability detection and SQL generation, with an abstention module implemented through confidence score thresholding and execution error filtering.

\subsection{Other Implementation Details}

Both CBR-to-SQL and RAG-to-SQL utilize Azure OpenAI LLMs: GPT-4o \citep{openai2024gpt4ocard} for main evaluation and the less capable GPT-4.1-mini \citep{openai2025gpt4.1} for ablation studies. The models are restricted to publicly available dataset-specific question–SQL pairs (MIMICSQL and EHRSQL) and decontextualized entities from the underlying EHR databases (MIMIC-III and MIMIC-IV Demo), ensuring no patient-identifiable information is exposed. Retrieval is implemented using a local Qdrant vector database \citep{qdrant2025}, which allows efficient nearest-neighbor search and supports metadata for question–SQL linking. For SQL execution and EHR storage, we utilize an in-memory SQLite \citep{sqlite} instance, ensuring that sensitive data is stored only for the session's duration. 

Unlike MIMICSQL, EHRSQL requires models to detect and abstain from answering impossible questions. To implement this, we augment the final SQL generation stage of both CBR-to-SQL and RAG-to-SQL with a prompt instruction to output ``None" when insufficient evidence for answerability is provided by the retrieved examples and schema descriptions, serving as a simple abstention mechanism.

\section{Results}

\subsection{Complete Database Experiment}

\tableref{tab:ex-lf-results} reports performance in the CDB setup, where models are evaluated under the standard setting with abundant training examples. On MIMICSQL, CBR-to-SQL achieves state-of-the-art $Acc_{LF}$ and outperforms RAG-to-SQL on both metrics, demonstrating that its structure-focused retrieval produces higher-quality few-shot examples than single-step retrieval. When compared with fine-tuned methods, CBR-to-SQL achieves comparable $Acc_{EX}$ to MedTS (0.899 vs. 0.894), but MedTS's large $Acc_{EX}$-to-$Acc_{LF}$ gap suggests far weaker logical structure generalization, which is notable given MIMICSQL's relatively simple SQL structures. The approach that clearly surpasses CBR-to-SQL on $Acc_{EX}$ is \citet{CHEN2026111800}, which augments retrieval with a heterogeneous knowledge graph providing substantially richer context, though the absence of $Acc_{LF}$ limits direct comparison to our method. On EHRSQL, CBR-to-SQL's advantage over RAG-to-SQL generalizes to a considerably more challenging dataset, achieving the highest $Acc_{EX}$ among all compared methods. We note that competing methods on EHRSQL are optimized for a penalty-based variant of $Acc_{EX}$ that accounts for unanswerable questions \citep{lee-etal-2024-overview}, rather than standard $Acc_{EX}$ or $Acc_{LF}$, which limits direct comparison. Nevertheless, CBR-to-SQL's strong performance across both datasets demonstrates that it can exploit the data abundance more effectively than standard RAG.

\begin{table*}[h]
\centering
\small
\caption{Comparison of methods on $Acc_{EX}$ and $Acc_{LF}$ metrics in CDB. Higher metric values indicate stronger performance. Cells marked ``--'' indicate that the authors do not report the metric.}
\setlength{\tabcolsep}{3.5pt}
% ===== MIMICSQL =====
\begin{minipage}[t]{0.475\textwidth}
\centering
\textbf{(a) MIMICSQL}
\vspace{1mm}

\begin{tabular}{lcc}
\toprule
\textbf{Method} & $\mathbf{Acc_{EX}} (\uparrow)$ & $\mathbf{Acc_{LF}} (\uparrow)$ \\
\midrule
SQLNet        & 0.260 & 0.142 \\
PtrGen        & 0.292 & 0.180 \\
Coarse2Fine   & 0.378 & 0.496 \\
TREQS         & 0.654 & 0.556 \\
MedTS         & 0.899 & 0.784 \\
\cite{CHEN2026111800}        & \textbf{0.942} & -- \\
\midrule
RAG-to-SQL (Ours)    & 0.860 & 0.825 \\
CBR-to-SQL (Ours)    & 0.894 & \textbf{0.862} \\
\bottomrule
\end{tabular}
\end{minipage}
\hfill
% ===== EHRSQL =====
\begin{minipage}[t]{0.475\textwidth}
\centering
\textbf{(b) EHRSQL}
\vspace{1mm}
\begin{tabular}{lcc}
\toprule
\textbf{Method} & $\mathbf{Acc_{EX}} (\uparrow)$ & $\mathbf{Acc_{LF}} (\uparrow)$ \\
\midrule
LTRC-IIITH     & 0.668 &  -- \\
AIRI NLP       & 0.689 &  -- \\
KU-DMIS        & 0.721 &  -- \\
ProbGate       & 0.819 &  -- \\
PromptMind     & 0.826 &  -- \\
PLUQ & 0.882 &  -- \\
\midrule
RAG-to-SQL (Ours) & 0.818 & 0.676 \\
CBR-to-SQL (Ours) & \textbf{0.895} & \textbf{0.729} \\
\bottomrule
\end{tabular}
\end{minipage}
\label{tab:ex-lf-results}
\end{table*}

\tableref{tab:cdb-brittle-results} evaluate model brittleness, defined as the sensitivity of performance to the removal of top-ranked retrieved examples. Results show that CBR-to-SQL is consistently less brittle than RAG-to-SQL, exhibiting smaller performance drops on both metrics across datasets. Notably, both models experience larger drops on the more complex EHRSQL dataset, yet CBR-to-SQL maintains a significantly smaller decline in $Acc_{EX}$. This suggests that CBR-to-SQL does not rely heavily on a few specific high-ranking examples (exact matches) to generalize. Instead, it can effectively leverage abstract structural patterns from retrieved examples, improving adaptiveness to a wider range of queries.

\begin{table}[!h]
\centering
\small
\caption{Comparison of methods on brittleness metric in CDB. Lower $\Delta$ values indicate less brittle behavior.}
\setlength{\tabcolsep}{6pt} 
\begin{tabular}{lcccc}
\toprule
\multirow{2}{*}{\textbf{Method}} 
& \multicolumn{2}{c}{\textbf{MIMICSQL}} 
& \multicolumn{2}{c}{\textbf{EHRSQL}} \\
\cmidrule(lr){2-3} \cmidrule(lr){4-5}
& $\Delta_{brittle^{EX}} (\downarrow)$ & $\Delta_{brittle_{LF}} (\downarrow)$ 
& $\Delta_{brittle_{EX}} (\downarrow)$ & $\Delta_{brittle_{LF}} (\downarrow)$ \\
\midrule
RAG-to-SQL   & 0.074 & 0.052 & 0.138 & 0.149 \\
CBR-to-SQL   & \textbf{0.028} & \textbf{0.033} & \textbf{0.043} & \textbf{0.128} \\
\bottomrule
\end{tabular}
\label{tab:cdb-brittle-results}
\end{table}

\subsection{Incomplete Database Experiment}

\tableref{tab:idb-ex-lf-results} evaluates performance under IDB, where training data is reduced to below 10\% of its original size. CBR-to-SQL maintains its lead over RAG-to-SQL on both datasets, with the performance gap nearly doubling compared to CDB in most cases. This gap is notable on EHRSQL, where CBR-to-SQL outperforms RAG-to-SQL by 11.9\% on $Acc_{EX}$ and 10.2\% on $Acc_{LF}$. This indicates that under low example diversity, RAG-to-SQL's single-step retrieval struggles to find near-exact matches and fails to generalize, while CBR-to-SQL's abstract skeletons allow more reliable structural matching, and its entity alignment is resolved via a lookup table, making it more robust to changes in the knowledge base. Remarkably, CBR-to-SQL in IDB outperforms RAG-to-SQL in the CDB setup across nearly all metrics, demonstrating that CBR-to-SQL can match or exceed its baseline with less than 10\% of the training data. The only exception is EHRSQL's $Acc_{LF}$, where RAG-to-SQL in CDB maintains a slight edge, likely because $Acc_{LF}$ is heavily influenced by example diversity, especially for EHRSQL with more complex SQL queries. 

\begin{table}[!h]
\centering
\small
\caption{Comparison of methods on $Acc_{EX}$ and $Acc_{LF}$ metrics in IDB.}
\setlength{\tabcolsep}{6pt} 
\begin{tabular}{lcccc}
\toprule
\multirow{2}{*}{\textbf{Method}} 
& \multicolumn{2}{c}{\textbf{MIMICSQL}} 
& \multicolumn{2}{c}{\textbf{EHRSQL}} \\
\cmidrule(lr){2-3} \cmidrule(lr){4-5}
& $Acc_{EX} (\uparrow)$ & $Acc_{LF} (\uparrow)$ 
& $Acc_{EX} (\uparrow)$ & $Acc_{LF} (\uparrow)$ \\
\midrule
RAG-to-SQL   & 0.781 & 0.779 & 0.718 & 0.542 \\
CBR-to-SQL   & \textbf{0.871} & \textbf{0.836} & \textbf{0.837} & \textbf{0.644} \\
\bottomrule
\end{tabular}
\label{tab:idb-ex-lf-results}
\end{table}

The brittleness results in IDB (\tableref{tab:idb-brittleness}) show substantially larger performance drops than in CDB, highlighting the difficulty of this limited setting. Notably, in the more complex EHRSQL dataset, CBR-to-SQL exhibits a performance drop on $Acc_{LF}$ nearly as large as RAG-to-SQL. Despite this, CBR-to-SQL consistently exhibits smaller declines than its baseline across IDB and CDB setups. This consistency is meaningful, as it implies that CBR-to-SQL's robustness is not a result of data abundance but its ability to retrieve structurally relevant examples and resolve entities effectively, even under limited data availability.

\begin{table}[!h]
\centering
\small
\caption{Comparison of methods on brittleness metric in IDB.}
\setlength{\tabcolsep}{6pt}
\begin{tabular}{lcccc}
\toprule
\multirow{2}{*}{\textbf{Method}} 
& \multicolumn{2}{c}{\textbf{MIMICSQL}} 
& \multicolumn{2}{c}{\textbf{EHRSQL}} \\
\cmidrule(lr){2-3} \cmidrule(lr){4-5}
& $\Delta_{brittle_{EX}} (\downarrow)$ & $\Delta_{brittle_{LF}} (\downarrow)$ 
& $\Delta_{brittle_{EX}} (\downarrow)$ & $\Delta_{brittle_{LF}} (\downarrow)$ \\
\midrule
RAG-to-SQL   & 0.082 & 0.079 & 0.171 & 0.184 \\
CBR-to-SQL   & \textbf{0.065} & \textbf{0.061} & \textbf{0.086} & \textbf{0.182} \\
\bottomrule
\end{tabular}
\label{tab:idb-brittleness}
\end{table}

\subsection{Ablation Studies}

We analyze the contributions of individual CBR-to-SQL components within the CDB setup in \tableref{tab:abla-ex-lf-results}. Two ablations were tested: removing Revision, and replacing Proposal with standard RAG retrieval (i.e., no masking). Removing Revision causes a sharp performance drop, as expected: Proposal alone captures only abstract patterns, leaving the model to generate SQL without entity alignment. Interestingly, replacing Proposal with standard retrieval results in a smaller but notable decline. This suggests that without masking, single-step RAG retrieval suffers from noise and struggles to identify optimal task demonstrations. This finding implies that greater specificity is not inherently beneficial and validates the importance of focusing on the abstract problem structure during retrieval.

\begin{table}[!h]
\centering
\caption{$Acc_{EX}$ and $Acc_{LF}$ when each online component in CBR-to-SQL is ablated in CDB for MIMICSQL, compared to its original performance.}
\begin{tabular}{lcc}
\toprule
\textbf{Method} & $Acc_{EX} (\uparrow)$ & $Acc_{LF} (\uparrow)$ \\
\midrule
CBR-to-SQL & \textbf{0.894} & \textbf{0.862} \\
\midrule
Replace \textit{Proposal} stage & 0.889 & 0.851 \\
No \textit{Revision} stage & 0.786 & 0.776 \\
\bottomrule
\end{tabular}
\label{tab:abla-ex-lf-results}
\end{table}

\begin{figure}[!h]
\centering
\includegraphics[width=\linewidth, keepaspectratio]{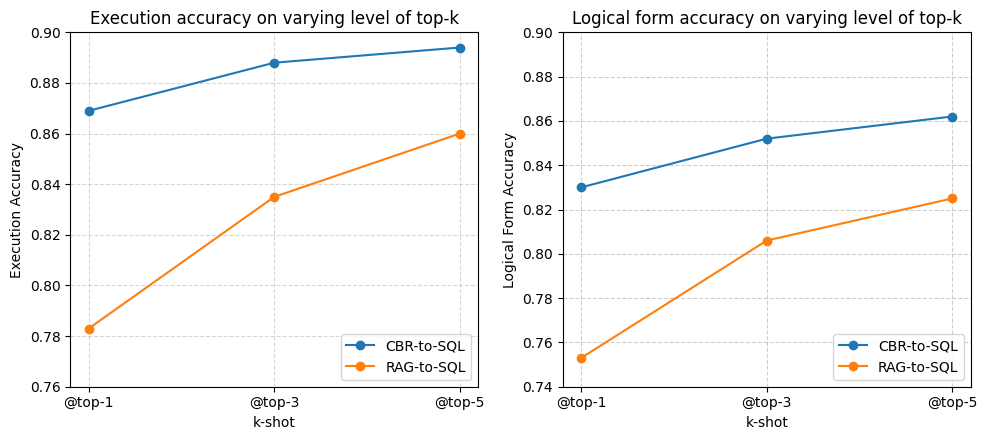}
\caption{$Acc_{EX}$ and $Acc_{LF}$ of different methods on different numbers of $k$-shot examples ($k\in\{1, 3, 5\}$) in CDB for MIMICSQL.}
\label{fig:topk-res-mimic}
\end{figure}

We additionally investigate the sensitivity of each system to the number of retrieved examples ($k$) in \figureref{fig:topk-res-mimic}. Overall, CBR-to-SQL consistently outperforms RAG-to-SQL across all values of $k$ in the CDB setup and its performance also declines more gracefully than the baseline as $k$ decreases. The performance gap is most notable for $Acc_{EX}$ and $Acc_{LF}$ in the one-shot scenario ($k=1$), where models receive the minimum amount of context. This demonstrates CBR-to-SQL's robustness in low-context settings where standard retrieval-based approaches inherently struggle.

Additional experiments on model time and token consumption, RAG enhancement techniques integration, fine-grained component accuracy, and qualitative error analysis are provided in \appendixref{apd:add-studies}.

\section{Discussion} 

\label{sec:discussion}

\paragraph{Architectural Advantages} 

Using the same underlying LLM, our experiments demonstrate that CBR-to-SQL's performance advantage over RAG-to-SQL stems from its architecture design. While RAG utilizes single-step retrieval, CBR-to-SQL decomposes the problems of retrieving logical structures and resolving entity references into two explicit stages, allowing each stage to optimize for its specific sub-problem independently. Further ablation studies (\appendixref{tab:abla-lf-breakdown-results}) confirm each component's intended role: Proposal provides reusable structural examples, and Revision aligns entities to their schema-accurate values. For clinical applications, this separation of responsibilities reduces the risk of black-box incorrect queries by allowing each component to fail transparently and independently.

Furthermore, as demonstrated in \appendixref{apd:rag-based-enhancements}, any advanced retrieval technique (hybrid retrieval, neural re-ranking) compatible with RAG can equally extend CBR-to-SQL. This follows naturally from CBR-to-SQL's modular design, where retrieval and entity alignment are decoupled into separate stages, allowing retrieval enhancements to be applied to the Proposal stage without altering the broader framework. This modularity ensures that with CBR-to-SQL, clinical deployments are not ``locked" into a single retrieval strategy and can directly benefit from ongoing advances in the broader RAG ecosystem.

Finally, CBR-to-SQL allows the two components to draw from separate data sources: the Proposal stage learns logical patterns from the knowledge base of question–SQL pairs, while the Revision stage aligns entities directly to the EHR database via the lookup table. This modularity makes CBR-to-SQL more adaptable, as abstracted example skeletons can be curated and improved independently of the underlying database schema. This reduces maintenance overhead and allows continuous improvement of each component without the costly retraining of the entire system.

\paragraph{Sample Efficiency}

From \tableref{tab:ex-lf-results} and \ref{tab:idb-ex-lf-results}, we observe that CBR-to-SQL trained in the IDB setup, which is less than 10\% of the complete data, outperforms RAG-to-SQL in CDB, demonstrating significantly higher sample efficiency. Furthermore, in $k$-shot ablation tests (\figureref{fig:topk-res-mimic}), CBR-to-SQL consistently outperforms RAG-to-SQL across all values of $k$ and exhibits more robust degradation as $k$ decreases. These results demonstrate that CBR-to-SQL can extract knowledge from available examples more effectively than standard RAG, allowing CBR-to-SQL to exploit the abundance of data in CDB while maintaining robustness when knowledge is scarce in IDB. From a clinical perspective, this means that institutions with limited labeled data can deploy reliable text-to-SQL systems, while those with extensive data can achieve even higher accuracy.

\paragraph{Interpretability} 

CBR-to-SQL's multi-stage architecture provides interpretability into the reasoning process. With RAG's single-step nature, it can be unclear to diagnose how the retrieval process fails to identify correct candidates. In contrast, by separating the tasks of structure retrieval and entity alignment, the stages are more transparent and their failures can be easily traceable, as evidenced in the error analysis in \appendixref{apd:error-analysis}. This transparency is particularly important in real-world deployment, where clinicians need to understand not just \textit{whether} a query failed but \textit{why} it failed, allowing them to isolate the failing component and apply targeted fixes accordingly.

\paragraph{Computational Cost} 

From the efficiency study in \appendixref{apd:efficiency}, the trade-off of higher performance is higher computational cost, as using CBR-to-SQL yields more time and token usage compared to RAG-to-SQL. However, the marginal cost analysis in \tableref{tab:marginal-cost} shows that these increases are modest compared to the gains in $Acc_{LF}$ and especially $Acc_{EX}$. Importantly, this calculation has not considered the hidden long-term costs of using RAG approaches to match the performance of CBR-to-SQL, such as maintaining a significantly larger knowledge base and managing higher error rates. Furthermore, in real-world healthcare settings, the advantages of accuracy, robustness, and scalability of CBR-to-SQL can outweigh its additional cost, particularly when considering that even small improvements in query accuracy can reduce the risk of incorrect EHR insights and flawed clinical decisions.

\section{Limitations \& Future Work}

This study has several limitations. First, many system components are not yet fully optimized. For instance, qualitative error analysis in \appendixref{apd:error-analysis} shows that entity tagging errors frequently propagate to downstream modules. Future work can address this with a feedback-based entity tagging that continuously refines extracted text spans and their tags using an entity lookup table, correcting tagging errors early in the pipeline. In addition, the simplistic lookup table design in Revision can be substituted with a more advanced medical terminology search system in practical deployments. 

Second, computational constraints limited our experiments to a proof-of-concept setup, preventing extensive exploration of alternative LLMs and hyperparameter settings. Future work could extend in this direction by testing CBR-to-SQL on other open-source and privately hosted LLMs to assess its generalization across different LLM backends. In addition, we have not explored the \textit{continual learning} aspect of CBR, where successful and failed inferences could be incorporated back into the knowledge base for ongoing improvement. This idea of continual knowledge improvement is one of the main principles of CBR theory \citep{aamodt1994case} and can be a highly promising direction for future work.

Finally, our evaluation is limited to EHR schemas in the medical domain. Although the MIMIC-* variants provide a realistic clinical benchmark, it does not reflect the diversity of schemas in other domains. As a result, the generalizability of CBR-to-SQL to broader text-to-SQL settings remains an open question. Future work should extend CBR-to-SQL to open-domain benchmarks like Spider \citep{yu-etal-2018-spider} or WikiSQL \citep{zhong2017seq2sqlgeneratingstructuredqueries} to assess its ability to handle more diverse text-to-SQL structures and entity types.

% ACKNOWLEDGEMENTS ONLY GO IN THE CAMERA-READY, NOT THE SUBMISSION
% \acks{Many thanks to all collaborators and funders!}

%Do NOT change font size of references or modify the bibliography style
\bibliography{sample}

\newpage
\appendix

\section{Case-based Reasoning Cycle}
\label{apd:cbr-cycle}

\begin{figure}[!h]
    \centering
    \includegraphics[width=350pt, keepaspectratio]{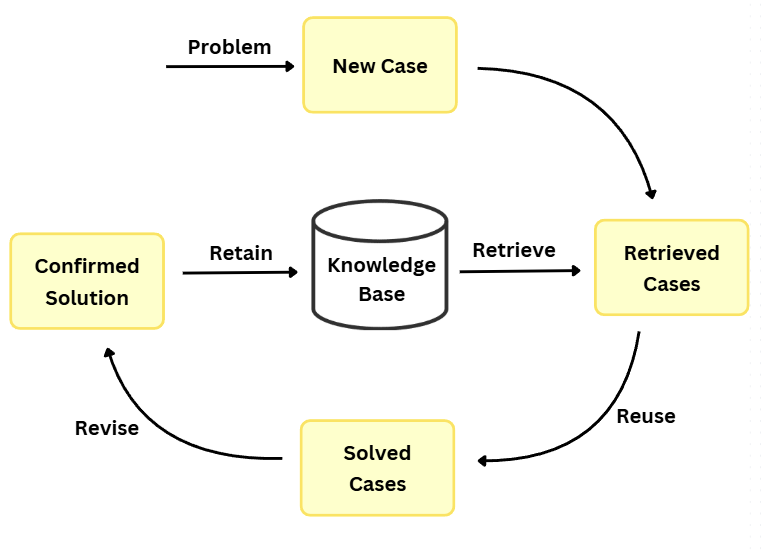}
    \caption{CBR cycle adapted from \cite{aamodt1994case}.}
    \label{fig:cbr-cycle}
\end{figure}

As illustrated in \figureref{fig:cbr-cycle}, CBR is conceptually described in a 4-stage cycle (the four RE-stages): REtrieve, REuse, REvise, and REtain. The \textit{retrieval} process is guided by a similarity metric to assess how closely a past case aligns with the new problem, allowing the retrieval of relevant problem-solving patterns. Once relevant cases are obtained, their solutions are \textit{reused}, often requiring further \textit{revision} to adapt to the existing problem. Finally, the knowledge base is enriched by \textit{retaining} the solved cases for future reference. This cycle allows CBR systems to continuously accumulate new experiences and incrementally refine their problem-solving strategies over time. 

\section{Lookup Table Construction}
\label{apd:entity-categories}

\subsection{Automatic Column Filtering}

The lookup table maps entity mentions to their database locations, but only for entities that require disambiguation. We therefore limit its contents to name-driven, semantically meaningful text values that exhibit terminological and semantic variation, excluding identifiers, codes, and numerical data that contribute little to entity alignment. This is achieved through an automated filtering process that identifies and retains only columns containing meaningful entities within the database schema. The lookup table construction script was run independently on the training splits of MIMICSQL and EHRSQL, producing a dataset-specific lookup table for each dataset.

The filtering process begins with a simple regular expression filter to exclude columns containing purely numerical data, UUIDs, or random alphanumeric strings. Columns that pass this initial screening proceed to a semantic richness evaluation, which assesses whether the column contains meaningful, name-driven textual values. This evaluation utilizes an LLM judge that scores each column based on ten randomly sampled entities, considering clinical and linguistic context. The semantic richness scoring scheme is as follows:

\begin{itemize}
    \item \textbf{Rich text (0.7–1.0)} includes name-driven, clinically meaningful values such as multi-word medical terms (e.g., ``Diabetes Mellitus Type 2", ``Acute Renal Failure"), full descriptive names (e.g., ``Emergency Department", ``Intensive Care Unit"), and complete clinical terms (e.g., ``Metformin", ``Elective").
    \item \textbf{Moderate text (0.4–0.7)} includes short but meaningful entities, including medical abbreviations with context (e.g., ``ICU", ``CABG"), single meaningful words (e.g., ``Male", "Hispanic"), language codes in appropriate columns (e.g., ``ENGL" in a language column), and status terms (e.g., ``URGENT").
    \item \textbf{Low-scoring text (0.0–0.3)} includes values that are not semantically useful, such as pure numbers, random identifiers (e.g., ``ABC123XYZ"), meaningless codes, or single characters lacking interpretable meaning. Locational context is explicitly considered: for instance, ``ENGL" receives a moderate score (0.5) in a language column, while the same value in a different context might score lower.
\end{itemize}

Columns exceeding a configurable threshold (in this case, 0.5) are retained, and all distinct values from these columns are extracted, preprocessed (de-duplication, case normalization), and indexed alongside their source table and column as metadata. The resulting lookup table allows efficient retrieval of relevant database locations given an entity mention during inference, allowing similar entity retrieval in the presence of typos, abbreviations, or non-standard terminology.

\subsection{Column Filtering Results}

As described in \tableref{tab:lookup-stats} and \tableref{tab:column-examples}, the filtering process effectively retained semantically rich columns while discarding non-informative identifiers, codes, and numerical data. For MIMICSQL, 19 of the 70 total columns (27\%) passed both filters, yielding 82,685 entities. For EHRSQL, 26 of 226 columns (12\%) were retained, providing 200,146 entities. The values of the included columns span concepts such as patient demographics, diagnoses, laboratory tests, prescriptions, and procedures. The disparity in retention rates between datasets reflects differences in schema design: MIMICSQL contains more pre-processed, human-readable text columns (e.g., DIAGNOSIS, LONG\_TITLE), while EHRSQL includes numerous clinical event tables with primarily numeric or coded values. The resulting lookup tables provide comprehensive coverage of medical terminology for downstream entity alignment tasks.

\begin{table}[h]
\centering
\caption{Lookup table construction statistics.}
\label{tab:lookup-stats}
\begin{tabular}{lcc}
\toprule
\textbf{Metric} & \textbf{MIMICSQL} & \textbf{EHRSQL} \\
\midrule
Tables Processed      & 5     & 17    \\
Columns Analyzed      & 70    & 226   \\
Text Columns          & 30    & 97    \\
Passed Simple Filter  & 22    & 75    \\
Passed LLM Filter     & 19    & 26    \\
Entities Indexed      & 82,685 & 200,146 \\
\bottomrule
\end{tabular}
\end{table}

\begin{table}[h]
\centering
\caption{Accepted and rejected column examples by category.}
\label{tab:column-examples}
\begin{tabular}{lll}
\toprule
\textbf{Category} & \textbf{MIMICSQL} & \textbf{EHRSQL} \\
\midrule
\multicolumn{3}{l}{\textbf{Accepted Columns}} \\
\midrule
Demographics & \texttt{NAME, MARITAL\_STATUS,} & \texttt{admission\_type,} \\
             & \texttt{RELIGION, ETHNICITY} & \texttt{admission\_location,} \\
             & & \texttt{discharge\_location,} \\
             & & \texttt{insurance} \\
Diagnoses    & \texttt{SHORT\_TITLE, LONG\_TITLE} & \texttt{long\_title} \\
             & & \texttt{(d\_icd\_diagnoses)} \\
Procedures   & \texttt{SHORT\_TITLE, LONG\_TITLE} & \texttt{long\_title} \\
             & & \texttt{(d\_icd\_procedures)} \\
Laboratory   & \texttt{LABEL, FLUID, CATEGORY} & \texttt{label (d\_labitems),} \\
             & & \texttt{valueuom, fluid} \\
Medications  & \texttt{DRUG, DRUG\_TYPE} & \texttt{drug, dose\_unit\_rx,} \\
             & & \texttt{route} \\
Clinical units & - & \texttt{first\_careunit,} \\
             &   & \texttt{last\_careunit,} \\
             &   & \texttt{careunit, eventtype} \\
\midrule
\multicolumn{3}{l}{\textbf{Rejected Columns}} \\
\midrule
Numeric IDs  & \texttt{SUBJECT\_ID, HADM\_ID} & \texttt{subject\_id, hadm\_id,} \\
             & & \texttt{row\_id} \\
Codes        & \texttt{ICD9\_CODE} & \texttt{icd\_code, itemid} \\
Timestamps   & \texttt{ADMITTIME, DISCHTIME,} & \texttt{admittime, dischtime,} \\
             & \texttt{DOB} & \texttt{charttime} \\
Measurements & - & \texttt{valuenum,} \\
             &   & \texttt{dose\_val\_rx} \\
Flags        & - & \texttt{flag} \\
\bottomrule
\end{tabular}
\end{table}

\section{Dataset Statistics}

\subsection{MIMICSQL}
\label{apd:mimicsql-statistics}

As shown in \tableref{tab:mimicsql-stats}, the human-modified set of MIMICSQL contains 10,000 question-SQL pairs, covering a relational schema spanning over 40,000 patients in five different tables, where their detailed descriptions are given in \tableref{tab:mimic_tables}. 

\begin{table*}[h]
\centering
\caption{Statistics of the MIMICSQL dataset \citep{wang2020texttosqlgenerationquestionanswering}.}
\begin{tabular}{ll}
\toprule
\textbf{Statistic} & \textbf{Value} \\
\midrule
Number of patients & 46,520 \\
Number of tables & 5 \\
Number of columns in tables \footnote{Columns are listed in order: Demographics, Diagnoses, Procedures, Prescriptions, Laboratory Tests.} & 23 / 5 / 5 / 7 / 9 \\
Number of Question-SQL pairs & 10,000 \\
Average template question length (words) & 18.39 \\
Average natural language question length (words) & 16.45 \\
Average SQL query length (tokens) & 21.14 \\
Average number of aggregation columns & 1.10 \\
Average number of conditions & 1.76 \\
\bottomrule
\end{tabular}
\label{tab:mimicsql-stats}
\end{table*}

\begin{table*}[h]
\centering
\caption{Overview of tables in MIMICSQL \citep{wang2020texttosqlgenerationquestionanswering}.}
\begin{tabular}{lp{10cm}}
\toprule
\textbf{Table}      & \textbf{Description} \\ 
\midrule
Demographics        & Contains patient demographic information such as age, gender, ethnicity, admission details, and other identifiers. \\
Diagnoses           & Records all diagnosis codes (ICD-9) assigned during hospital stays, capturing the medical conditions and comorbidities of patients. \\
Procedures          & Contains procedural codes (ICD-9) for interventions or surgeries performed on patients during their hospital stay. \\
Prescriptions       & Details of medication orders including drug name, dosage, route, frequency, and timing, prescribed to patients. \\ 
Laboratory Tests    & Laboratory measurements and test results such as blood tests, biochemistry, and hematology with timestamps. \\
\bottomrule
\end{tabular}
\label{tab:mimic_tables}
\end{table*} 

\subsection{EHRSQL}
\label{apd:ehrsql-statistics}

EHRSQL covers large relational schema of 26 tables, covering 100 patients. The statistics of data splits are provided in \tableref{tab:ehrsql_statistics}. More details related to the schema tables can be found in \cite{PhysioNet-mimic-iv-demo-2.2} and \cite{PhysioNet-mimiciv-3.1}.

\begin{table*}[h]
\centering
\caption{Data statistics for the EHRSQL shared task \citep{lee-etal-2024-overview}.}
\begin{tabular}{lccc}
\hline
 & \textbf{Train} & \textbf{Valid} & \textbf{Test} \\
\hline
Answerable question template 
& 100 
& 134 
& 134 \\

Answerable samples 
& 4674 
& 931 
& 934 \\

Unanswerable samples 
& 450 
& 232 
& 233 \\

\hline
Total samples 
& 5124 
& 1163 
& 1167 \\
\hline
\end{tabular}
\label{tab:ehrsql_statistics}
\end{table*}

\section{IDB Construction}
\label{apd:idb-construction}

\subsection{Clustering Process}

The masked SQL embeddings are clustered using HDBSCAN \citep{hdbscan}, an unsupervised algorithm capable of detecting clusters of varying shapes and densities without requiring a predefined number of clusters. Unlike DBSCAN \citep{10.5555/3001460.3001507}, HDBSCAN builds a hierarchy of density-based clusters and selects the most stable ones, allowing it to capture the natural grouping of the data while labeling low-density points as noise. In this context, the primary advantage of HDBSCAN is its ability to automatically determine the number and shape of clusters. Detecting noise is rather a secondary concern, since masked SQL queries are already standardized and naturally can form well-defined clusters.

\begin{figure}
    \centering
    \includegraphics[width=\linewidth, keepaspectratio]{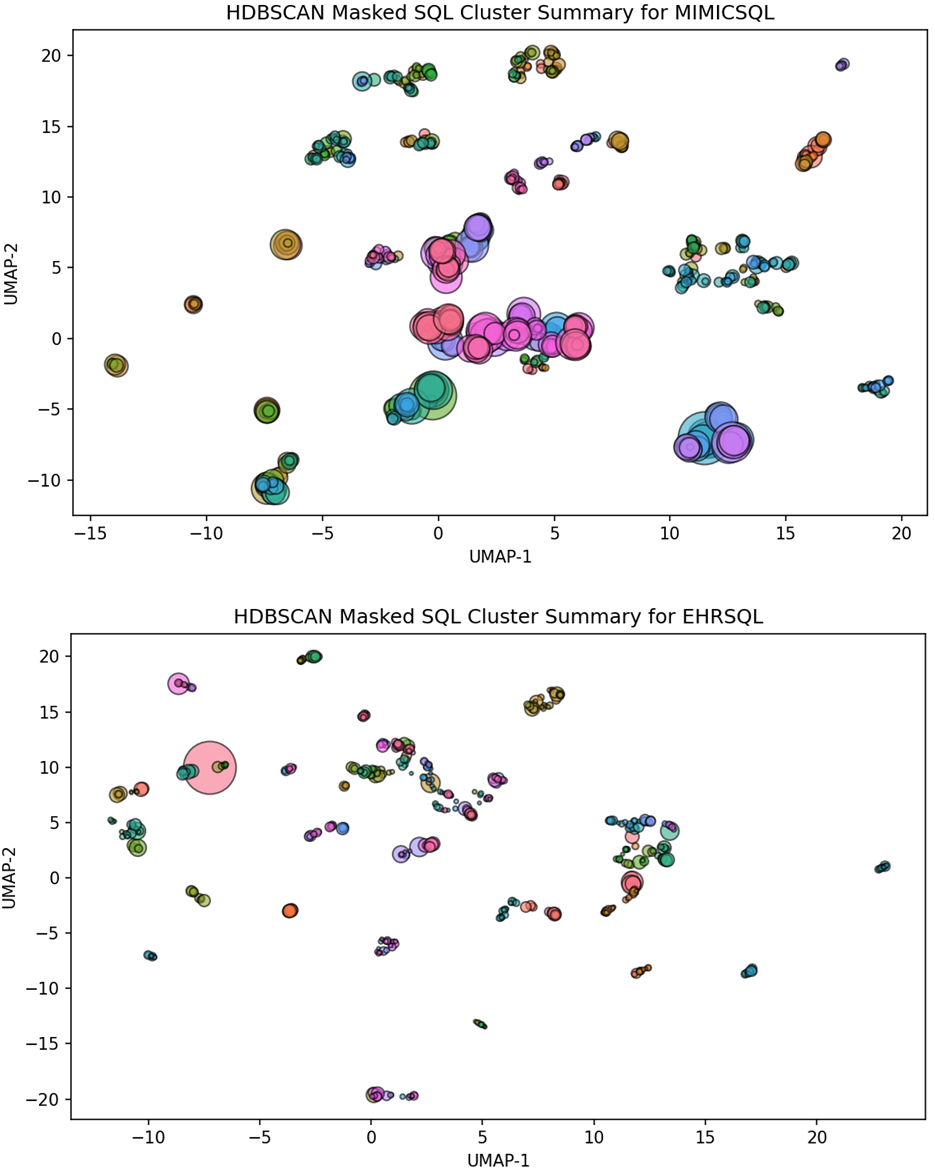}
    \caption{UMAP projection of the HBSCAN clustering results over masked SQL queries in the IDB setup. Each circle represents a cluster, and the circle size reflects the cluster size.}
    \label{fig:cluster-result}
\end{figure}  

HDBSCAN's clustering process is configured with a minimum cluster size of 2 and a tight distance threshold $\epsilon$ of 0.10, encouraging small, highly compact, yet well-defined groupings. This ensures that the extracted examples represent concrete, specific patterns rather than overly abstract generalizations. Overall, for MIMICSQL, the algorithm identified 679 clusters, with 95 datapoints (question-SQL pairs) labeled as noise (i.e., not assigned to any cluster). For EHRSQL, it identified 475 clusters, with 23 noise datapoints. A large cluster consisting entirely of impossible questions, where the corresponding SQL is always ``NULL", was discarded from the IDB retaining process, as these queries do not contribute meaningful structural patterns. Selecting one representative datapoint from each cluster and including all noise datapoints yielded a total of 774 unique datapoints for MIMICSQL and 498 datapoints for EHRSQL. Relative to the original dataset sizes, approximately 8,000 datapoints for MIMICSQL and 5,124 for EHRSQL; this represents a substantial reduction, preserving less than 10\% of the original data.

\subsection{Clustering Results}

\figureref{fig:cluster-result} visualizes the clustering results for both datasets. Overall, MIMICSQL exhibits more larger clusters, while EHRSQL consists primarily of smaller, more fragmented clusters. This difference is expected, as SQL queries in EHRSQL are generally more complex and varied, naturally leading to smaller, more specialized groupings. The one abnormally large cluster in EHRSQL corresponds to the impossible questions in the training data, which will be discarded prior to ingestion to IDB.

\tableref{tab:cluster_questions} shows three randomly selected natural language questions in the largest clusters. Overall, it is possible to observe that the questions are highly similar in intent and structure across clusters, demonstrating that SQL-level clustering effectively captures logical and structural patterns. 

\begin{table*}[!h]
\centering
\caption{Representative questions from the largest HDBSCAN clusters in the Incomplete Database (IDB) setup across MIMICSQL and EHRSQL datasets. The clustering effectively groups structurally similar SQL queries.}
\begin{tabular}{l l p{0.65\textwidth}}
\toprule
\textbf{Dataset} & \textbf{Cluster Size} & \textbf{Representative Questions} \\
\midrule

MIMICSQL & \textbf{124} & 
\parbox[t]{\hsize}{%
How many patients were diagnosed with anxiety state NOS and were given drugs orally?\\
What is the number of patients whose diagnoses title is acute nephritis nec and drug route is po?\\
Provide the number of patients who are diagnosed with Wegener's granulomatosis and have sc route of drug administration.
} \\
\addlinespace

 & \textbf{102} & 
\parbox[t]{\hsize}{%
Count the number of patients whose primary disease is left internal jugular vein thrombosis; left arm edema and who died in or before the year 2115.\\
How many patients died of colangitis in or before the year 2180?\\
What is the number of patients who had rash as primary disease and died in 2155 or before that?
} \\
\addlinespace

\midrule

EHRSQL & \textbf{75} & 
\parbox[t]{\hsize}{%
Tell me the price of trimethoprim.\\
Prazosin – How much does it cost?\\
What are the prices of the drug albumin 25\% (12.5g / 50ml)?
} \\
\addlinespace

 & \textbf{72} & 
\parbox[t]{\hsize}{%
How much does the pH change in patient 10021666 last measured on the last hospital visit compared to the first value measured on the last hospital visit?\\
What are patient 10027602's changes in glucose, CSF second measured on the first hospital visit compared to the first value measured on the first hospital visit?\\
How much difference is there in phosphate levels among patient 10015931 second measured on the first hospital visit compared to the value first measured on the first hospital visit?
} \\
\addlinespace

\bottomrule
\end{tabular}
\label{tab:cluster_questions}
\end{table*}

\section{Baseline Architecture}
\label{apd:ragsql-architecture}

\begin{figure*}
\centering
\includegraphics[width=380pt, keepaspectratio]{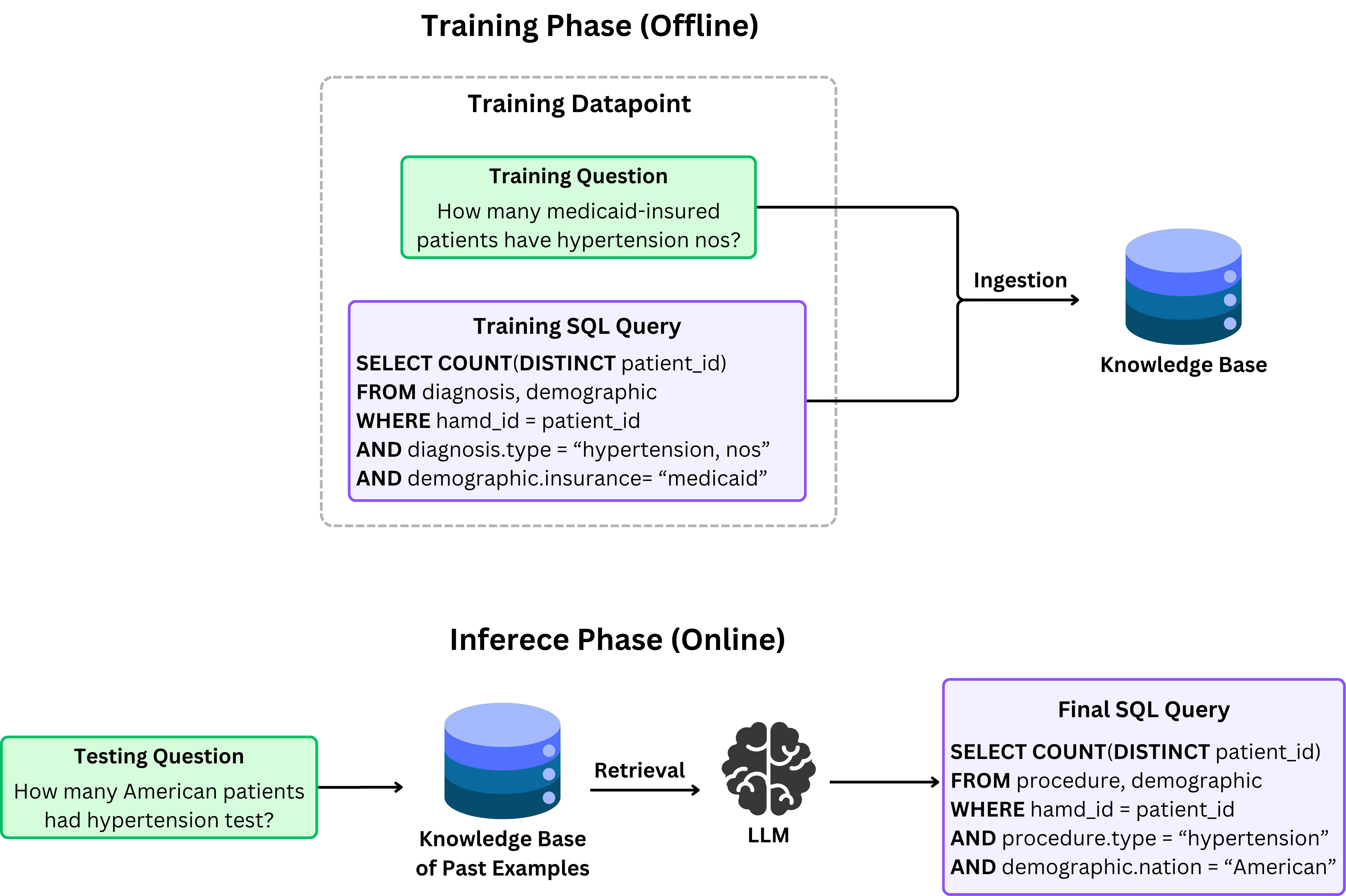}
\caption{Overview of RAG-to-SQL's architecture.}
\label{fig:rag2sql-architecture}
\end{figure*}

The detailed RAG-to-SQL architecture is given in \figureref{fig:rag2sql-architecture}.

\section{Additional Studies}
\label{apd:add-studies}

\subsection{Efficiency Aspects}
\label{apd:efficiency}

\tableref{tab:efficiency} reports the average latency and token cost per run for CBR-to-SQL and its baseline in the CDB setup. As expected, RAG-to-SQL has notably lower inference time and token consumption. This efficiency stems from its simpler architecture, as retrieved examples are directly appended to the LLM without additional processing steps. In contrast, CBR-to-SQL's higher cost is a result of its multi-stage design, which requires generating intermediate representations and processing longer prompts during online inference.

\begin{table}[!h]
\centering
\small
\caption{Average latency (in seconds) and token usage per query across two datasets in CDB.}
\setlength{\tabcolsep}{6pt}
\begin{tabular}{lcccc}
\toprule
\multirow{2}{*}{\textbf{Method}} 
& \multicolumn{2}{c}{\textbf{MIMICSQL}} 
& \multicolumn{2}{c}{\textbf{EHRSQL}} \\
\cmidrule(lr){2-3} \cmidrule(lr){4-5}
& Latency ($\downarrow$) & Token ($\downarrow$) 
& Latency ($\downarrow$) & Token ($\downarrow$) \\
\midrule
RAG-to-SQL 
& \textbf{1.67} & \textbf{2251.68} 
& \textbf{3.02} & \textbf{6819.99} \\
CBR-to-SQL 
& 2.60 & 5399.09 
& 4.43 & 9512.79 \\
\bottomrule
\end{tabular}
\label{tab:efficiency}
\end{table}

To better understand the efficiency-performance trade-off, \tableref{tab:marginal-cost} presents the marginal cost per metric gain for $Acc_{EX}$ and $Acc_{LF}$ between the two models. The values are obtained by dividing the average performance difference in the metric score ($Acc_{EX}$ or $Acc_{LF}$) by the cost difference (token usage or latency). The calculations are based on IDB results rather than CDB, since the tighter constraints on available demonstrations in IDB make accuracy improvements more meaningful. Overall, the results suggest that each 1\% increase in execution accuracy comes at a cost of roughly 0.1 seconds and 200-300 tokens, which are within practical bounds for real-world applications, especially in clinical settings where correctness outweighs efficiency. While improvements in logical form accuracy are more expensive, this is expected, as matching the exact SQL structure ground-truth answer is generally more demanding, requiring more instruction prompts or a larger example coverage.

\begin{table}[!h]
\centering
\small
\caption{Marginal cost of performance gain for CBR-to-SQL over RAG-to-SQL across two datasets in IDB.}
\setlength{\tabcolsep}{6pt}
\begin{tabular}{lcccc}
\toprule
\multirow{2}{*}{\textbf{Metric}} 
& \multicolumn{2}{c}{\textbf{MIMICSQL}} 
& \multicolumn{2}{c}{\textbf{EHRSQL}} \\
\cmidrule(lr){2-3} \cmidrule(lr){4-5}
& \makecell{Per 1\% \\ $Acc_{EX}$ Gain} 
& \makecell{Per 1\% \\ $Acc_{LF}$ Gain}
& \makecell{Per 1\% \\ $Acc_{EX}$ Gain} 
& \makecell{Per 1\% \\ $Acc_{LF}$ Gain} \\
\midrule
Latency Overhead ($\downarrow$) 
& 0.103 & 0.163 
& 0.118 & 0.138 \\
Token Overhead ($\downarrow$) 
& 349.71 & 552.18 
& 226.29 & 264.00 \\
\bottomrule
\end{tabular}
\label{tab:marginal-cost}
\end{table}

\subsection{RAG-based Enhancements}
\label{apd:rag-based-enhancements}

In this experiment, we show that any retrieval enhancement compatible with RAG can equally extend CBR-to-SQL. We evaluate two such techniques: \textit{neural re-ranking} and \textit{hybrid retrieval}. Neural re-ranking introduces a cross-encoder model after the initial retrieval step, which scores each candidate by jointly encoding it with the query rather than relying on independent vector similarity, producing a more accurate relevance ranking at the cost of additional inference time. On the other hand, hybrid retrieval combines dense vector search with sparse keyword-based search (BM25), expanding candidate coverage by capturing both semantic and lexical similarity. For the implementation of neural re-ranking, we first retrieve an initial set of $k=15$, then apply a cross-encoder re-ranker to the unmasked questions to narrow the results down to 5. For hybrid retrieval, we augment both dense retrieval methods in CBR-to-SQL and RAG-to-SQL with a separate sparse BM25 stream over the unmasked questions to allow keyword matching. The results are then merged and re-ranked to $k=5$ datapoints using reciprocal rank fusion \citep{rrf}. For our evaluation setting, we select the EHRSQL with the CDB setup, as its more complex SQL structures leave more room to observe improvements from these RAG-based enhancements.

\tableref{tab:rag-enhancements} reports $Acc_{EX}$ and $Acc_{LF}$ on EHRSQL for both methods under each enhancement. Overall, both systems improve over their respective baselines when enhancements are applied. While there seems to be diminishing gains from CBR-to-SQL with these enhancements compared to RAG, this can be explained by CBR-to-SQL already reaching its performance ceiling on the dataset. Nevertheless, CBR-to-SQL with enhancements consistently outperforms its equivalent baselines across settings. The most substantial gains come from hybrid retrieval: CBR-to-SQL + Hybrid Retrieval achieves the highest $Acc_{EX}$ (0.909) and $Acc_{LF}$ (0.775), improving over the base CBR-to-SQL by 1.4\% and 4.6\% respectively. Neural re-ranking yields more modest gains for CBR-to-SQL, with a marginal drop in $Acc_{EX}$ and a smaller improvement in $Acc_{LF}$ compared to hybrid retrieval, suggesting that cross-encoder re-ranking provides limited additional signal on top of CBR-to-SQL's already structure-focused retrieval.  

Even though we observed higher performance for CBR-to-SQL with these techniques, the improvements are not without trade-offs. For instance, neural re-ranking produces a marginal drop in $Acc_{EX}$ for CBR-to-SQL, suggesting that utilizing these retrieval heuristics does not always translate to uniform improvements across metrics. In any case, enhancements can trade off loss against gains on another, and whether they can yield desired improvements depends heavily on the specific setting. Nevertheless, the results support the view that CBR-to-SQL can be effectively combined with other RAG-based enhancements.

\begin{table}[t]
\centering
\small
\caption{Ablation study of RAG-based enhancements on EHRSQL in CDB, applied to both RAG-to-SQL and CBR-to-SQL.}
\setlength{\tabcolsep}{3.5pt}
\begin{tabular}{lcc}
\toprule
\textbf{Method} & $\mathbf{Acc_{EX}} (\uparrow)$ & $\mathbf{Acc_{LF}} (\uparrow)$ \\
\midrule
RAG-to-SQL                        & 0.818 & 0.677 \\
RAG-to-SQL + Neural Re-ranking    & 0.828 & 0.725 \\
RAG-to-SQL + Hybrid Retrieval     & 0.842 & 0.733 \\
\midrule
CBR-to-SQL                        & 0.895 & 0.729 \\
CBR-to-SQL + Neural Re-ranking    & 0.893 & 0.739 \\
CBR-to-SQL + Hybrid Retrieval     & \textbf{0.909} & \textbf{0.775} \\
\bottomrule
\end{tabular}
\label{tab:rag-enhancements}
\end{table}

\subsection{Fine-grained Clause-level Accuracy Analysis}
\label{apd:breakdown-accuracy}

We present the breakdown of $Acc_{LF}$ across different syntactic aspects of SQL queries. For this, the authors of MIMICSQL, \cite{wang2020texttosqlgenerationquestionanswering}, defined the aspects according to the following SQL template:
\begin{equation}
\begin{aligned}
\label{eq:sql-template}
&\text{SELECT } Agg_{op} (Agg_{col})^+ \\
&\text{FROM } Table \\
&\text{WHERE } (Con_{col}\ Con_{op}\ Con_{val})^+,
\end{aligned}
\end{equation}
where the superscript $+$ indicates one or more items.

In this template, $Agg_{op}$ is the operation used for the selected column $Agg_{col}$ and can be the following values: “COUNT”, “MAX”, “MIN”, “AVG”, and finally “NULL”, representing no operation. $Agg_{col}$ is the name of the selected column to be transformed using $Agg_{op}$ to answer the user question. $Table$ represents table references, which can be a join of different tables. The section in the WHERE clause contains the filtering conditions, and each condition takes the form of $(Con_{col}\ Con_{op}\ Con_{val})$, where $Con_{col}$ is the selected column, $Con_{op}$ is the filtering operation (``$=$'', ``$>$'', ``$<$'', ``$\le$'', ``$\ge$''), and $Con_{val}$ represents the compared value. Overall, $Agg_{col}$ and $Agg_{op}$ reflect the model’s ability to interpret user intent by correctly mapping the input question to the appropriate aggregation columns and operations, while $Table$, $Con_{col+op}$, and $Con_{val}$ reflect the model’s ability to identify target entities, resolve their locations within the schema, and generate correct condition clauses with them.

\paragraph{CDB Environment}

From \tableref{tab:breakdown-lf-results}, retrieval-based methods (RAG-to-SQL and CBR-to-SQL) excel at entity linking ($Con_{val}$), with CBR-to-SQL achieving the highest score. This reflects the ability of retrievers to obtain examples with relevant entities, allowing the LLM to predict the correct condition value accurately. However, RAG-to-SQL and CBR-to-SQL perform worse on $Agg_{col}$ than fine-tuned approaches, reflecting weaker aggregation column prediction. This may be explained by the fact that fine-tuning allows MedTS to better internalize frequent question–schema associations during training, while retrieval-based models are constrained by the limited diversity of retrieved examples ($k=5$), which may not always contain column selections that align closely with the input query.

Despite this, CBR-to-SQL shows a more balanced score range across aspects compared to RAG-to-SQL. More importantly, it consistently outperforms RAG-to-SQL across every metric on both datasets. The most notable gains come from $Con_{val}$ and $Con_{col+op}$, the components most dependent on accurate entity alignment. On MIMICSQL, CBR-to-SQL improves $Con_{val}$ by 2.9\% and $Con_{col+op}$ by 2.3\% over RAG-to-SQL. On the more challenging EHRSQL dataset, the performance gap widens further: $Con_{val}$ improves by 4.0\% and $Con_{col+op}$ by 5.0\%. The consistently strong gains on condition-related components confirm that by separating structure and entity tasks, CBR-to-SQL can achieve more accurate alignment without sacrificing structural alignment.

\begin{table*}[t]
\centering
\caption{$Acc_{LF}$ of clause-specific breakdown matching in CDB.}
% ===== Block (a) Baselines =====
\begin{minipage}[t]{\textwidth}
\centering
\textbf{(a) MIMICSQL}
\vspace{1mm}

\begin{tabular}{lccccc}
\toprule
\textbf{Method} & $Agg_{op} (\uparrow)$ & $Agg_{col} (\uparrow)$ & $Table (\uparrow)$ & $Con_{col+op} (\uparrow)$ & $Con_{val} (\uparrow)$ \\
\midrule
Coarse2Fine & 0.321 & 0.313 & 0.321 & 0.260 & 0.214 \\
Seq2Seq     & 0.978 & 0.872 & 0.926 & 0.471 & 0.174 \\
SQLNet      & \textbf{0.994} & 0.939 & 0.933 & 0.722 & 0.080 \\
PtrGen      & 0.987 & 0.917 & 0.944 & 0.795 & 0.236 \\
TREQS       & 0.990 & 0.912 & 0.942 & 0.834 & 0.694 \\
MedTS       & \textbf{0.994} & \textbf{0.988} & 0.971 & 0.893 & 0.785 \\
\midrule
RAG-to-SQL (Ours) & 0.993 & 0.947 & 0.972 & 0.904 & 0.871 \\
CBR-to-SQL (Ours) & \textbf{0.994} & 0.957 & \textbf{0.985} & \textbf{0.927} & \textbf{0.900} \\
\bottomrule
\end{tabular}
\end{minipage}

\vspace{6mm}

% ===== Block (b) Ours =====
\begin{minipage}[t]{\textwidth}
\centering
\textbf{(b) EHRSQL}
\vspace{1mm}

\begin{tabular}{lccccc}
\toprule
\textbf{Method} & $Agg_{op} (\uparrow)$ & $Agg_{col} (\uparrow)$ & $Table (\uparrow)$ & $Con_{col+op} (\uparrow)$ & $Con_{val} (\uparrow)$ \\
\midrule
RAG-to-SQL & 0.931 & 0.779 & 0.831 & 0.740 & 0.712 \\
CBR-to-SQL & \textbf{0.952} & \textbf{0.872} & \textbf{0.905} & \textbf{0.790} & \textbf{0.752} \\
\bottomrule
\end{tabular}
\end{minipage}
\label{tab:breakdown-lf-results}
\end{table*}

\paragraph{IDB Environment}

From \tableref{tab:idb-lf-breakdown-results}, CBR-to-SQL outperforms or matches RAG-to-SQL across most aspects in the limited IDB setting. The largest improvements once again occur in $Con_{val}$ and $Con_{col+op}$, the components most dependent on accurate entity alignment and semantic understanding of conditions. On MIMICSQL, CBR-to-SQL improves $Con_{val}$ by 5.3\% and $Con_{col+op}$ by 3.5\%. On EHRSQL, the gains are even more notable, as $Con_{val}$ improves by 9.9\% and $Con_{col+op}$ by 9.9\%. This considerably larger performance gap under data scarcity showcases CBR-to-SQL's ability to extract meaningful patterns from limited examples, whereas RAG-to-SQL struggles when fewer relevant examples are available for retrieval.

On MIMICSQL, RAG-to-SQL achieves a slightly higher $Agg_{op}$, though the difference is negligible and both models operate near ceiling on this component. This likely reflects a trade-off to CBR-to-SQL's design, as the masking step that strips away surface details to improve entity alignment can occasionally remove cues useful for predicting simple aggregation operators. Meanwhile, RAG-to-SQL preserves all surface information and can copy operators directly from retrieved examples. Nevertheless, the fact that the difference is very minimal (0.1\%) and both scores are near-perfect suggests this is not a meaningful limitation.

\begin{table*}[t]
\centering
\caption{$Acc_{LF}$ of clause-specific breakdown matching in IDB.}
% ===== Block (a) Baselines =====
\begin{minipage}[t]{\textwidth}
\centering
\textbf{(a) MIMICSQL}
\vspace{1mm}

\begin{tabular}{lccccc}
\toprule
\textbf{Method} & $Agg_{op} (\uparrow)$ & $Agg_{col} (\uparrow)$ & $Table (\uparrow)$ & $Con_{col+op} (\uparrow)$ & $Con_{val} (\uparrow)$ \\
\midrule
RAG-to-SQL & \textbf{0.993} & 0.943 & 0.948 & 0.875 & 0.830 \\
CBR-to-SQL & 0.992 & \textbf{0.944} & \textbf{0.980} & \textbf{0.910} & \textbf{0.883} \\
\bottomrule
\end{tabular}
\end{minipage}

\vspace{6mm}

% ===== Block (b) Ours =====
\begin{minipage}[t]{\textwidth}
\centering
\textbf{(b) EHRSQL}
\vspace{1mm}

\begin{tabular}{lccccc}
\toprule
\textbf{Method} & $Agg_{op} (\uparrow)$ & $Agg_{col} (\uparrow)$ & $Table (\uparrow)$ & $Con_{col+op} (\uparrow)$ & $Con_{val} (\uparrow)$ \\
\midrule
RAG-to-SQL & 0.909 & 0.670 & 0.738 & 0.622 & 0.580 \\
CBR-to-SQL & \textbf{0.947} & \textbf{0.791} & \textbf{0.857} & \textbf{0.721} & \textbf{0.679} \\
\bottomrule
\end{tabular}
\end{minipage}
\label{tab:idb-lf-breakdown-results}
\end{table*}

\subsection{Further Ablation Studies}
\label{apd:ablation}

We assessed the impact of the underlying language model by evaluating both methods using the less capable GPT-4.1-mini in \tableref{tab:mini-ex-lf-results}. Despite reduced LLM capability, CBR-to-SQL retains a consistent performance advantage (+2.1\% $Acc_{EX}$, +1.5\% $Acc_{LF}$) over the baseline and even outperforms RAG-to-SQL with GPT-4o in MIMICSQL's CDB setting. These results indicate that the observed performance gains primarily derive from the architectural design rather than the underlying LLM.

\begin{table}[!h]
\centering
\caption{Comparison of methods on $Acc_{EX}$ and $Acc_{LF}$ metrics in CDB for MIMICSQL using GPT-4.1-mini as the underlying LLM.}
\begin{tabular}{lcc}
\toprule
\textbf{Method} & $Acc_{EX} (\uparrow)$ & $Acc_{LF} (\uparrow)$ \\
\midrule
RAG-to-SQL & 0.841 & 0.820 \\
CBR-to-SQL & \textbf{0.862} & \textbf{0.835} \\
\bottomrule
\end{tabular}
\label{tab:mini-ex-lf-results}
\end{table}

% We further examined the fine-grained contributions of individual system components using clause-level accuracy. From \tableref{tab:abla-lf-breakdown-results}, removing Revision produces the steepest decline, causing misalignment with elements connected to the entities, thus resulting in large drops in $Con_{val}$, $Con_{col+op}$, and $Table$. In contrast, other aspects remain stable, as Revision does not directly affect aggregation structure.

% Replacing Proposal with standard retrieval causes only minor overall losses. Interestingly, $Con_{col+op}$ and $Con_{val}$ rise slightly, since the RAG-based retriever can exploit more specific examples. However, this comes with the trade-off of increased noise, leading to notable declines in $Agg_{col}$ (–1.6\%) and $Agg_{op}$ (–1.0\%), reflecting the importance of example masking for retrieving structurally aligned examples.

We further examined the fine-grained contributions of individual system components using clause-level accuracy. From \tableref{tab:abla-lf-breakdown-results}, removing Revision produces the steepest overall decline, with accuracy dropping sharply on aspects tied to entity disambiguation: $Table$ falls 3.8\%, $Con_{col+op}$ drops 6.0\%, and $Con_{val}$ declines by 8.4\%. This is expected, as Revision is responsible for resolving ambiguous entity mentions to their correct schema locations. In contrast, aggregation-related components ($Agg_{op}$, $Agg_{col}$) remain stable, since Revision does not affect the query's high-level structure.

Replacing Proposal with standard RAG-based retrieval yields a more mixed picture. Interestingly, this ablated model achieves slightly higher $Agg_{col}$ and maintains comparable $Agg_{op}$, suggesting that the RAG retriever can better retrieve examples with similar aggregation structure. However, this comes at a trade-off of increased noise, causing notable drops on other aspects: $Table$ at –1.4\%, $Con_{col+op}$ at -1.8\%, and $Con_{val}$ at -1.1\%. This indicates that example masking plays an important role in structural alignment and is essential for retrieving structurally relevant examples during the Proposal stage.

\begin{table}[h!]
\centering
\caption{Logical form accuracy of break-down matching in CDB setup for MIMICSQL, in ablation settings.}
\begin{tabular}{lccccc}
\toprule
\textbf{Method} & $Agg_{op} (\uparrow)$ & $Agg_{col (\uparrow)}$ & $Table (\uparrow)$ & $Con_{col+op} (\uparrow)$ & $Con_{val} (\uparrow)$ \\
\midrule
CBR-to-SQL & \textbf{0.994} & 0.957 & \textbf{0.985} & \textbf{0.927} & \textbf{0.900} \\
\midrule
Replace \textit{Proposal} & 0.993 & \textbf{0.958} & 0.971 & 0.909 & 0.889 \\
No \textit{Revision} & \textbf{0.994} & 0.952 & 0.947 & 0.867 & 0.816 \\
\bottomrule
\end{tabular}
\label{tab:abla-lf-breakdown-results}
\end{table}

\subsection{Qualitative Error Analysis}
\label{apd:error-analysis}

To investigate the root causes of CBR-to-SQL failures, the study conducts a qualitative error analysis, focusing on cases where CBR-to-SQL fails but RAG-to-SQL succeeds, using $Acc_{EX}$ as the evaluation metric. Four commonly observed error categories are identified and presented in \tableref{tab:qual_cbr_failures}. Note that this analysis is limited to MIMICSQL in the CDB setting, which is our primary benchmark.

\textit{Canonical Form Mismatch} occurs when CBR-to-SQL finds multiple plausible matches for an entity but selects the incorrect one. For instance, ``guillain barre syndrome'' appears in both form in the table as ``GUILLAIN-BARRE SYNDROME" and ``GUILLAIN BARRE SYNDROME". Without further context, CBR-to-SQL selects the incorrect variant, while RAG-to-SQL can perform exact matching against retrieved examples to identify the correct entity form. This error arises from noisy data during lookup table ingestion, reflecting real-world database inconsistencies and highlighting the difficulty of maintaining clean entity representations in practice. This could be mitigated by introducing more thorough entity deduplication during lookup table construction to consolidate variant forms into one canonical version before ingestion.

\textit{Incorrect Source Retrieval} occurs when Revision fails to retrieve the correct entity. For example, ``CSF'' is detected, but the lookup table misses ``Cerebrospinal Fluid (CSF)'' and instead returns superficially similar but irrelevant entries like ``Glucose, CSF'' or ``WBC, CSF''. This stems from the use of Levenshtein distance, which does not account for medical abbreviations, acronyms, or aliases. Improving this step may require specialized mechanisms such as introducing domain-specific aliases for each entity or a specialized medical entity search engine.

\textit{Unrecognized Entity} refers to cases when the entity tagging process fails to recognize information that should be masked away. In the case given by the example, CBR-to-SQL fails to realize that ``replaced'' is referring to a specific route of drug and therefore should be tagged and masked away. Without the entity extracted, the Revision process never attempts to resolve it, resulting in the incorrect prediction of the entity (``replaced'') as the fallback. In contrast, RAG-to-SQL can retrieve an exact example containing the correct form ``replace'' and copy it directly. This highlights a limitation of CBR-to-SQL, as it is overdependent on the extraction module to identify entities, so any tagging error can easily propagate through the pipeline and affect downstream stages.

Similarly, \textit{Incorrect Entity Tagging} occurs when a single entity is mistakenly split into two during tagging, causing Revision to generate separate condition values in the SQL query. While the split may follow natural language structure, it ignores schema constraints, highlighting that entity tagging lacks schema awareness and can propagate errors to downstream stages. A possible solution is to introduce iterative refinement, where entity matches retrieved from the database are used to correct and improve the initial tagging until a satisfactory match is found.

Despite the failures, we can clearly observe that by splitting one retrieval problem into explicit stages, CBR-to-SQL allows each error category to be directly traceable to a specific component: incorrect source retrieval stems from Levenshtein distance, entity tagging errors propagate from the extraction module, and canonical form mismatching by lack of preprocessing in the lookup table construction. In contrast, standard RAG offers little of such component-level attribution. Its single-step retrieval provides only a binary outcome (success, failure) without any intermediate indicators to diagnose which sub-task has failed.

\onecolumn
\begin{longtable}{p{1cm}p{13cm}}
\caption{Error categories where CBR-to-SQL consistently fails but RAG-to-SQL succeeds in MIMICSQL's CDB. Each category is illustrated by a representative example. The underlined text highlights the specific part of the problem that CBR-to-SQL fails to address.}
\label{tab:qual_cbr_failures} \\
\toprule
\textbf{Error} & \textbf{Representative Example} \\
\midrule
\endfirsthead

\multicolumn{2}{c}%
{{\bfseries \tablename\ \thetable{} -- continued from previous page}} \\
\toprule
\textbf{Error} & \textbf{Representative Example} \\
\midrule
\endhead

\midrule
\multicolumn{2}{r}{{Continued on next page}} \\
\endfoot

\bottomrule
\endlastfoot

\smash{\rotatebox[origin=r]{90}{Canonical Form Mismatch}} &  
\textbf{Question:} How many patients born before the year 2109 were primarily diagnosed for \ul{guillain barre syndrome}?
\medskip

\textbf{CBR-to-SQL (Fail):} 
\smallskip

\begin{minipage}{\linewidth}
\ttfamily
\raggedright
SELECT COUNT ( DISTINCT DEMOGRAPHIC."SUBJECT\_ID" ) \\
FROM DEMOGRAPHIC \\
WHERE \ul{DEMOGRAPHIC."DIAGNOSIS" = "GUILLAIN-BARRE SYNDROME"} \\
\hspace{1em} AND DEMOGRAPHIC."DOB\_YEAR" < "2109"
\end{minipage}
\medskip

\textbf{RAG-to-SQL (Success):} 
\smallskip

\begin{minipage}{\linewidth}
\ttfamily
\raggedright
SELECT COUNT ( DISTINCT DEMOGRAPHIC."SUBJECT\_ID" ) \\
FROM DEMOGRAPHIC \\
WHERE \ul{DEMOGRAPHIC."DIAGNOSIS" = "GUILLAIN BARRE SYNDROME"} \\
\hspace{1em} AND DEMOGRAPHIC."DOB\_YEAR" < "2109"
\end{minipage} \\
\addlinespace\midrule

\smash{\rotatebox[origin=r]{90}{Incorrect Source Retrieval}} & 
\textbf{Question:} Count the number of patients for whom \ul{csf lab test} was ordered.
\medskip

\textbf{CBR-to-SQL (Fail):} 
\smallskip

\begin{minipage}{\linewidth}
\ttfamily
\raggedright
SELECT COUNT ( DISTINCT DEMOGRAPHIC."SUBJECT\_ID" )\\
FROM DEMOGRAPHIC \\
INNER JOIN LAB \\
\hspace{1em} on DEMOGRAPHIC.HADM\_ID = LAB.HADM\_ID \\
WHERE \ul{LAB."LABEL" = "Glucose, CSF"}
\end{minipage}
\medskip

\textbf{RAG-to-SQL (Success):} 
\smallskip

\begin{minipage}{\linewidth}
\ttfamily
\raggedright
SELECT COUNT ( DISTINCT DEMOGRAPHIC."SUBJECT\_ID" ) \\
FROM DEMOGRAPHIC \\
INNER JOIN LAB \\
\hspace{1em} on DEMOGRAPHIC.HADM\_ID = LAB.HADM\_ID \\
WHERE \ul{LAB."FLUID" = "Cerebrospinal Fluid (CSF)"}
\end{minipage} \\
\addlinespace
\midrule

\smash{\rotatebox[origin=r]{90}{Unrecognized Entity}} & 
\textbf{Question:} What is the number of patients with \ul{replaced} route of drug administration?
\medskip

\textbf{CBR-to-SQL (Fail):} 
\smallskip

\begin{minipage}{\linewidth}
\ttfamily
\raggedright
SELECT COUNT ( DISTINCT DEMOGRAPHIC."SUBJECT\_ID" ) \\
FROM DEMOGRAPHIC \\
INNER JOIN PRESCRIPTIONS \\
\hspace{1em} ON DEMOGRAPHIC.HADM\_ID = PRESCRIPTIONS.HADM\_ID \\
WHERE \ul{PRESCRIPTIONS."ROUTE" = "REPLACED"}
\end{minipage}
\medskip

\textbf{RAG-to-SQL (Success):} 
\smallskip

\begin{minipage}{\linewidth}
\ttfamily
\raggedright
SELECT COUNT ( DISTINCT DEMOGRAPHIC."SUBJECT\_ID" ) \\
FROM DEMOGRAPHIC \\
INNER JOIN PRESCRIPTIONS \\
\hspace{1em} ON DEMOGRAPHIC.HADM\_ID = PRESCRIPTIONS.HADM\_ID \\
WHERE \ul{PRESCRIPTIONS."ROUTE" = "REPLACE"}
\end{minipage} \\
\addlinespace
\midrule

\smash{\rotatebox[origin=r]{90}{Incorrect Entity Tagging}} & 
\textbf{Question:} What is the maximum number of days of hospital stay for patients with \ul{hypertension but not coronary artery disease/cardiac cath}?
\medskip

\textbf{CBR-to-SQL (Fail):} 
\smallskip

\begin{minipage}{\linewidth}
\ttfamily
\raggedright
SELECT MAX ( DEMOGRAPHIC."DAYS\_STAY" ) \\
FROM DEMOGRAPHIC \\
WHERE \ul{DEMOGRAPHIC."DIAGNOSIS" = "HYPERTENSION"} \\
\hspace{1em} AND \ul{DEMOGRAPHIC."DIAGNOSIS" <> "CORONARY ARTERY DISEASE\textbackslash CARDIAC CATH"}
\end{minipage}
\medskip

\textbf{RAG-to-SQL (Success):} 
\smallskip

\begin{minipage}{\linewidth}
\ttfamily
\raggedright
SELECT MAX ( DEMOGRAPHIC."DAYS\_STAY" ) \\
FROM DEMOGRAPHIC \\
WHERE \ul{DEMOGRAPHIC."DIAGNOSIS" = "HYPERTENSION;RULE OUT CORONARY ARTERY DISEASE\textbackslash CARDIAC CATH"}
\end{minipage} \\
\addlinespace
\end{longtable}

\end{document}